\documentclass[useAMS,usenatbib]{mn2e}
\usepackage{color}
\usepackage{colortbl}
\usepackage{multirow}
\usepackage{dcolumn}
\usepackage{amsmath}
\usepackage{amssymb}
\usepackage{graphicx}
\usepackage{epsfig}
\usepackage{changebar}
\usepackage{natbib}
\usepackage{multirow}
\usepackage{subfigure}
\usepackage{txfonts}
\usepackage{relsize}
\usepackage{verbatim} 
\usepackage{natbib}
\newcommand {\NH}{NH$_3$}
\newcommand {\water}{H$_2$O}
\newcommand{\HII}{\mbox{HII}}
\newcommand{\HC}{\mbox{HC$_{3}$N}}
\newcommand{\ie}{i.e.,~}

\newcommand{\msun}{M$_\odot$}

\newcommand{\kms}{km~s$^{-1}$}
\newcommand{\cmthree}{cm$^{-3}$}
\newcommand{\cmtwo}{cm$^{-2}$}

\newcommand{\x}{$\times$}

\newcommand{\aap}{aap}
\newcommand{\araa}{araa}
\newcommand{\aaps}{aaps}
\newcommand{\mnras}{MNRAS}
\newcommand{\apj}{ApJ}
\newcommand{\apjl}{ApJl}
\newcommand{\pasp}{PASP}

\title[The G305 star forming complex: Wide-Area molecular mapping of \NH\ and \water\ masers]{The G305 star forming complex: Wide-Area molecular mapping of \NH\ and \water\ masers}
\author[L.Hindson]{L.Hindson$^{1,2}$\thanks{E-mail:
l.hindson@herts.ac.uk}, M.A. Thompson$^{1}$, J.S.  Urquhart$^{2}$, J.S. Clark$^{3}$, B. Davies$^{4,5}$\\
$^{1}$\rm Centre for Astrophysics Research, Science and Technology Research Institute, University of Hertfordshire, College Lane, Hatfield, AL10 9AB, UK\\ \rm $^{2}$  Australia Telescope National Facility, CSIRO Astronomy and Space Science, PO Box 76, Epping, NSW, 1710, Australia\\ 
$^{3}$\rm Department of Physics and Astronomy, The Open University, Walton Hall, Milton Keynes MK7 6AA\\
$^{4}$\rm Center for Imaging Science, Rochester Institute of Technology, 54 Lomb Memorial Drive, Rochester, NY 14623, USA\\
$^{5}$\rm School of Physics \& Astronomy, University of Leeds, Woodhouse Lane, Leeds LS2 9JT}

\begin{document}

\date{Accepted 07/06/2010. Received 19/05/2010 ; in original form }

\pagerange{\pageref{firstpage}--\pageref{lastpage}} \pubyear{2010}

\maketitle

\label{firstpage}
\begin{abstract}
We present wide area radio (12 mm) Mopra Telescope observations of the complex and rich massive star forming region G305. Our goals are to determine the reservoir for star formation within G305 using \NH\ to trace the dense molecular content, and thus, the gas available to form stars; estimate physical parameters of detected \NH\ clumps (temperature, column density, mass etc); locate current areas of active star formation via the presence of \water\ and methanol masers and the distribution of YSOs and ultra compact HII regions associated with this region.
This paper details the \NH\ (J,K), (1,1), (2,2) and (3,3) inversion transition and 22 GHz \water\ maser observations. We observed a $\sim$$1.5\degr \times 1\degr$ region with $\sim2$\arcmin\ angular resolution and a sensitivity of $\sim60$ mK per 0.4 \kms\ channel. 
We identify 15 \NH\ (1,1), 12 \NH\ (2,2) and 6 \NH\ (3,3) clumps surrounding the central \HII\ region. The sizes of the clumps vary between $<2.6$ and 10.1 pc, the average kinetic temperature of the gas is 25 K. We calculate clump masses of $>10^{4}$ \msun\ and find the total molecular mass of the complex to be $\sim6\times10^{5}$ \msun. We note the positions of 56 star formation tracers, and discover a high degree of correlation with detected \NH\ clumps. We have detected 16 \water\ masers, find they correlate with the detected ammonia clumps and in general are found closer to the \NH\ clump cores than star formation tracers of later evolutionary stages.

\end{abstract}


\section{Introduction}
High mass stars ($>8$ \msun) have a significant effect on their surrounding volume, injecting a large amount of energy into the natal cloud. This feedback energy takes the form of high velocity winds, outflows, expanding \HII\ regions, high energy radiation and eventually supernovae. These processes have a profound effect, potentially triggering new star forming events \citep{Elmegreen1977,Habing1979,Elmegreen2002} or alternatively disrupting future formation by dispersing the natal cloud \citep{Elmegreen1977}.

\begin{figure*}
\includegraphics[trim=0 50 0 0 ,width=0.75\linewidth,angle=270]{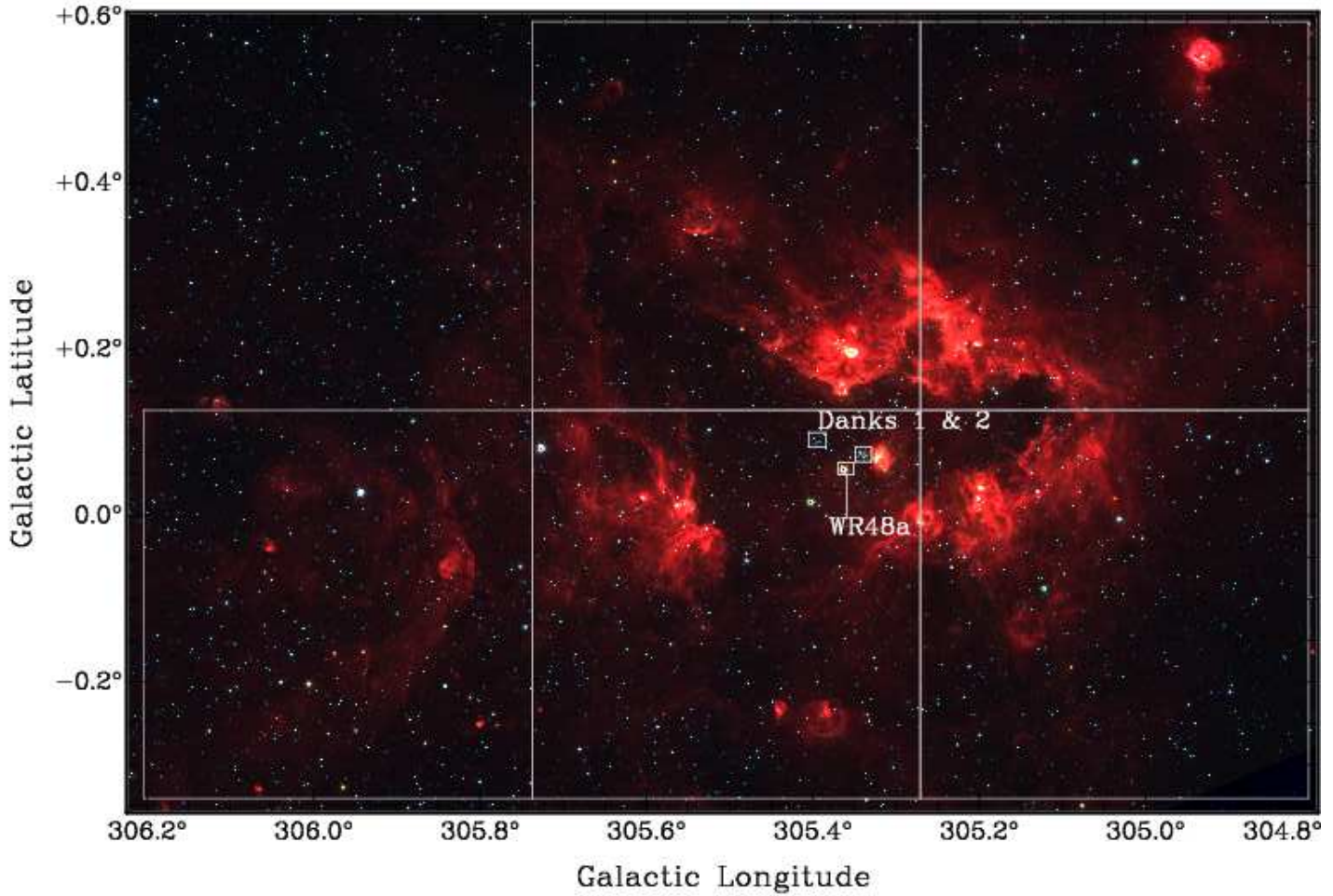}
\caption{Wide-field 3 colour image ( 8\,$\umu$m (red), 4.5\,$\umu$m (green) and 3.6\,$\umu$m (blue)) GLIMPSE image of the G305 complex, with the positions of Danks 1 \& 2 over-plotted with boxes. Our scan area is shown by light grey boxes.}
\label{G305 Glimpse 3 colour composite}
\end{figure*}

The G305 star forming region lies in the Scrutum Crux arm within the galactic plane at $l= 305\degr$, $b = 0\degr$. At a distance of $\sim 4$ kpc and with an age of 3-5 Myr, the complex is one of the most luminous \HII\ regions in the galaxy \citep{Clark2004}, centred on the open clusters Danks 1 \& 2 and the Wolf-Rayet star WR 48a. Numerous infrared hotspots, \HII\ emission, ultra compact (UC) \HII\ regions and molecular maser emission suggest that O type stars are currently forming, with a radio flux indicative of  $>31$ deeply embedded 07V stars \citep{Clark2004}. G305 provides an excellent laboratory in which to study massive star formation as it is one of the closest sites of massive star formation in the galaxy and contains multiple sites and different epochs of star formation over its large scale, allowing the investigation of evolution and environmental effects such as triggering.

In Fig. 1, we present a three colour composite image of G305, constructed using images obtained from the Galactic Legacy Infrared Mid-Plane Survey Extraordinaire (GLIMPSE) \citep{Benjamin2003,Churchwell2009}. The 8\,$\umu$m filter is of particular interest as it is dominated by polycyclic aromatic hydrocarbon (PAH) features, these are excited in the surface layers of molecular clouds that are exposed to a significant amount of UV radiation and therefore an excellent tracer of photo-dominated regions (PDRs) \citep{Leger1984}. For this reason PDRs are a good signpost for molecular regions that are undergoing a strong interaction with a nearby \HII\ region \citep{Urquhart2003}. In this image we show the positions of Danks 1 \& 2 and WR 48a which are believed to be the driving force behind the central diffuse \HII\ region \citep{Clark2004}.

Here we report on our first observations of \NH\ (1,1), (2,2), (3,3) and the 22 GHz \water\ maser line, using the Mopra radio telescope. Previous imaging of G305 by the \water\ Southern Galactic Plane Survey (HOPS) \citep{Walsh2008} has revealed the presence of \NH\ hotspots in G305, but lacked the depth to trace less massive clumps and moderate column densities. Using \NH\ which has an effective critical density of $10^4 - 10^5$ \cmthree\ \citep{Ho1983}, we are able to trace the dense gas structure of the region. Surveys of such dense regions are critical to relating the properties of the molecular gas to the star-forming properties of the cloud. We intend to make use of these low resolution \NH\ maps as a pathfinder for future high resolution interferometric observations. We also report the positions of 22 GHz \water\ maser emission, which have been shown to be good tracers of the early stages of low and high mass star formation \citep{Furuya2003}. In high mass YSOs, \water\ masers have been found to be associated with hot molecular cores, molecular outflows and jets. \water\ maser emission is believed to  highlight an evolutionary phase before the onset of a UC \HII\ region around an embedded star \citep{Furuya2003}. In low mass YSOs, \water\ masers are found to be associated with predominantly class 0 YSOs and in excited shocks of protostellar jets in the vicinity of the star. Thus, they are an excellent signpost of the early stages of star formation.

In Section 2 we present details of the observations and the data reduction processes, in Section 3 we present our observational results and derive basic parameters, Section 4 presents our discussion and in Section 5 we present a summary of our results.

\section{Observations \& data reduction procedures}

Observations were made using the Australia Telescope National Facility (ATNF) telescope Mopra. This is a 22m antenna located 26km outside the town of Coonabarrabran in New South Wales at an elevation of 866 meters above sea level and at a latitude of 31 degrees south.

The telescope is equipped  with a 12 mm receiver with a frequency range of 16 to 27.5 GHz. The UNSW Mopra spectrometer (MOPS) is made up of four 2.2 GHz bands which overlap slightly to provide 8 GHz continuous bandwidth. The Mopra spectrometer has two possible modes, narrow and broadband modes. The narrow band or ''zoom'' mode of MOPS was used and allowed us to observe 16 spectra simultaneously, with a bandwidth of 137.5 MHz over 4096 channels in each zoom window. This gives a channel spacing of 34 kHz, corresponding to a velocity resolution of $\sim0.4$ \kms\ at 24 GHz. We set our zoom windows to match those used in the HOPS survey, see Table 1 in \cite{Walsh2008} for details of targeted lines. Our line detections are summarised in Table 1. In this paper we focus upon the analysis of the \NH\ and \water\ maser detections. We  postpone the analysis of the remaining detections (H69$\alpha$, CH$_{3}$OH and HC$_{3}$N)  to future publications combining this data with forthcoming observations of the cm-wave radio continuum (Hindson et al 2010, in prep) and  multi-line studies of CH$_{3}$OH and \HC.

\begin{table}
\caption{Lines detected, central frequency and the size of the Mopra beam at the given frequency.}
  \begin{tabular}[h]{lcc}
   \hline Line & Freq & HPBW\\
           & (GHz) &(\arcsec) \\
 	\hline
\water\ (6-5)	&	22.235	&	144\\
\NH\ (1,1)		&	23.694	&	136\\
\NH\ (2,2)		&	23.722	&	136\\
\NH\ (3,3)		&	23.870	&	136\\
\HC\ (3,2)		&	27.294	&	120\\
H69$\alpha$		&	19.951	&	164\\
CH$_{3}$OH (3$_{2}$ - 3$_{1}$)	&	24.928	&	131\\
CH$_{3}$OH (4$_{2}$ - 4$_{1}$)	&	24.933	&	131\\
CH$_{3}$OH (2$_{2}$ - 2$_{1}$)	&	24.934	&	131\\	
CH$_{3}$OH (6$_{2}$ - 6$_{1}$)	&	24.959	&	131\\
CH$_{3}$OH (7$_{2}$ - 7$_{1}$)	&	25.018	&	131\\
\hline
\end{tabular}
\end{table}

In order to cover the  1.5 $\times$ 1 degree scale of G305 and remain within scan length constraints, our observations consisted of 5 $\times$ $28\arcmin\times28\arcmin$ maps with a $3\arcmin$ overlap between adjacent maps (see Fig. 1). All maps were obtained using the fly mapping mode, in which the telescope scans along the sky taking spectra at regular intervals along lines of constant latitude and longitude. The maps were sampled with $51\arcsec$ spacing to give Nyquist sampling at the highest frequency of 27.4 GHz and better than Nyquist at lower frequencies. The fastest scan rate of 2s per point was used, an off source position was observed periodically to remove sky emission. Combined with the limited scan speed and large map size a single map took $\sim 2$ hours to complete in one scan direction, and 4 hours with two orthogonal (Galactic latitude \& longitude) co added scans, used to reduce striping. We obtained 5 complete co added scans of each of the 5 regions, taking a total of 50 hours of observing time collected over a number of nights in April and May of 2009. The weather conditions were changeable over this period with water vapour levels between 15 and 25 mm and system temperatures of $\sim100$ K, with a variation in T$_{\rm sys}$ of no more than 20\% during the course of any particular map. During times of dense cloud cover and rain, observations were halted as these conditions caused system temperatures of well over 120 K.

The data were reduced using the packages LIVEDATA and GRIDZILLA, both are AIPS++ packages written by Mark Calabrreta\footnote{http://www.atnf.csiro.au/computing/software/livedata.html} for the Parkes radio telescope and adapted for Mopra. LIVEDATA performs a bandpass calibration for each row using the off-source data followed by fitting user specified polynomial to the spectral baseline. GRIDZILLA calculates the pixel value from the spectral values and weights using the weighted mean estimation. We use a Gaussian smoothing kernel of 2\arcmin\ with a cut off radius of $0.5\arcmin$. This produces a map with minimal smoothing giving a  pixel size of $0.5\arcmin\times 0.5\arcmin$ and a sensitivity of $\sim60$ mK per 0.4 \kms\ channel, approximately twice as deep as the ongoing HOPS survey \citep{Walsh2008}. We make use of a recent study into the efficiency of Mopra \citep{Urquhart2010} to convert the \NH\ antenna temperatures into main beam temperatures using the efficiency at 23 GHz of 0.64. 

\section{Results and analysis}

In this section, we describe the distribution of \NH\ clumps as well as the locations of detected \water\ masers in the G305 complex. We define individual \NH\ clumps using the clump finding algorithm {\sc Fellwalker} and present their source averaged spectra. We derive the basic physical properties of the detected clumps from their \NH\ emission.

\subsection{The distribution of ammonia clumps and \water\ masers}

In Fig. 2, we present peak temperature contour plots of \NH\ (1,1) and (2,2) emission, the noise in the map is $\sim 0.06$ K. We show the location of \water\ maser emission detected towards G305 (blue crosses). The ammonia contours are overlaid onto a grayscale GLIMPSE 5.4\,$\umu$m image, which also shows PAH emission highlighting the PDR. We use the 5.4\,$\umu$m image because it is not as bright as the 8\,$\umu$m band and so avoids the problem of saturation commonly encountered with the 8\,$\umu$m band. We detect a number of \NH\ clumps in four distinct regions around the central \HII\ region. Using WR 48a as a reference, we define the four regions of emission as (Galactic) north east (NE), north west (NW), south east (SE) and west (W). Within these regions, we distinguish 15 \NH\ (1,1), 12 \NH\ (2,2) clumps and 6 \NH\ (3,3) clumps. The majority of clumps are located around the periphery of the central cavity, with the higher transitions located mainly towards the PDR boundary. 
We detect 16 \water\ masers distributed throughout the region, all but two are found to be within the bounds of \NH\ emission. Of the 14 masers associated with \NH\ emission, eleven are found close to the clump cores, no \NH\ core is associated with more than one maser detection. Weak \NH\ clumps (excluding clump 11) in the western region appear to be devoid of \water\ maser emission.
Detected clumps have been defined and numbered according to the output generated by the clump finding algorithm {\sc Fellwalker} (see section 3.1.1) and \water\ masers are numbered according to Table 5.\\

\begin{figure*}
\caption{Contour map of the peak temperature  \NH\ (1,1) and (2,2) emission towards G305 in black and red contours respectively. Emission is  over-plotted onto a  GLIMPSE 5.4\,$\umu$m (greyscale) image, detected \water\ masers are shown by blue crosses and numbered according to Table 5. Contours begin at 0.15 K and increment by 0.1 K, the noise in the map is 0.06 K. The clump numbers are shown for reference within circles and clump areas are shown as light grey outlines. Following our nomenclature we separate the complex into four regions of emission with grey boxes.}
\includegraphics[width=0.6\textwidth,angle=270]{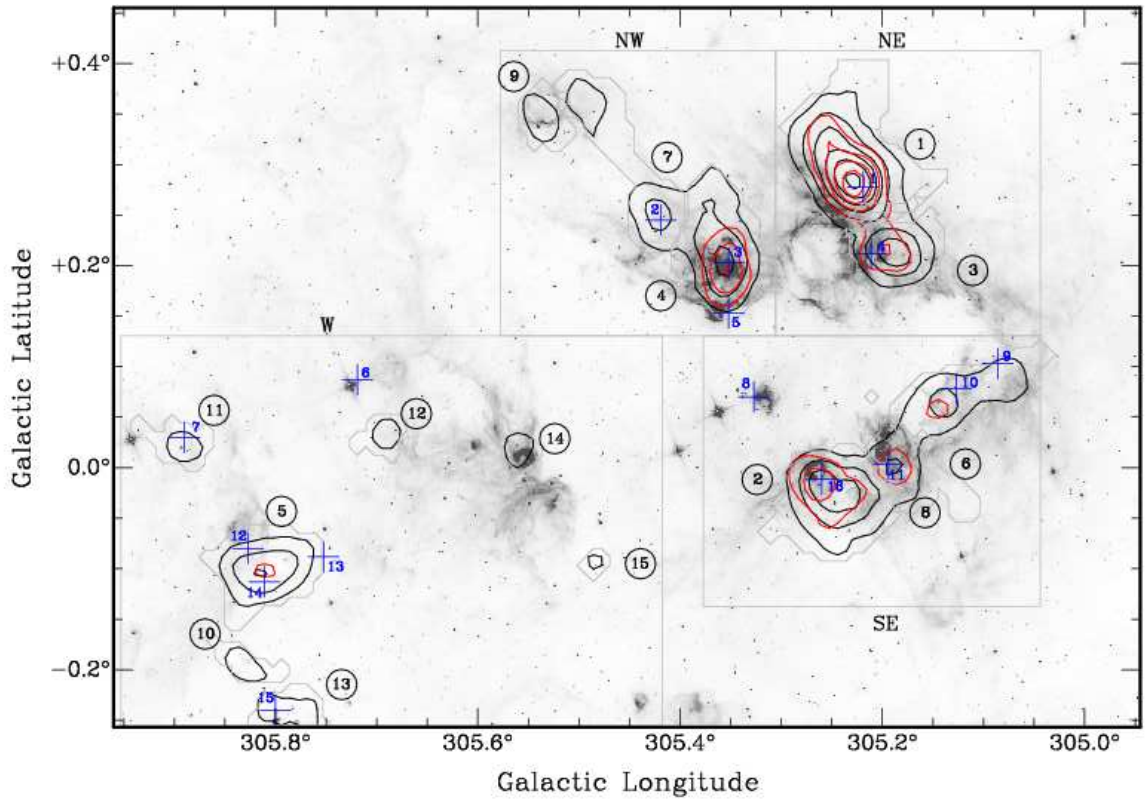}
\label{Main Results Map}
\end{figure*}

\noindent \textbf{NW Region:} Within this region we detect three \NH\ (1,1) clumps and one \NH\ (2,2) and (3,3) clump centred within clump 4. We detect three \water\ masers in this region within the boundaries of clump 4 and 7, these are close to but offset from the peak emission of the ammonia clumps. Ammonia clumps are connected by a surrounding region of low emission in the (1,1) transition and clump 4 appears to be elongated in the direction away from the central exciting stars. Unlike the other clumps in the complex, the emission peak in clump 4 is coincident with strong 5.4\,$\umu$m emission towards the centre, indicating that the direction of the powerful (possibly ionising) radiation is impacting from our perspective, face on with the cloud, this area of strong PAH emission is also coincident with a \HII\ (see \cite{Clark2004} Fig. 2). \\
 
\noindent \textbf{NE Region:} This region contains two \NH\ clumps positioned behind one another, with clump 1 positioned further to the north, both clumps exhibit \NH\ (1,1), (2,2) and (3,3) emission. We detect two \water\ masers in this region both are located near the cores of the detected clumps but again they are slightly offset from the \NH\ peak. Clump 1 is host to the strongest peak and average main beam temperatures in the complex in both \NH\ (1,1), (2,2) and (3,3) emission lines (1.21, 0.66 and 0.42 K respectively) and it is the largest single clump detected (10.1 pc). The clumps are  bounded by a region of low emission and are separated by a region of lower density. The two clumps are offset from what appears to be a bubble of intense PAH emission to the west that is also coincident with a \HII\ region (see \cite{Clark2004} Fig. 2).\\

\noindent \textbf{SE Region:} \NH\ (1,1) emission is observed in an unbroken strip parallel to the PDR and central cavity boundary, we observe three separate clumps within this strip, all of which are coincident with \NH\ (1,1) and (2,2) emission at the clump centres with \NH\ (3,3) detected within clump 2 and 8. We detect four water masers all within the bounds of the \NH\ (1,1) emission, three of these masers are found in close proximity to the detected ammonia clump peaks. Clump 2 appears to be extended in the direction away from the central clusters. We also note two strong PAH features on the periphery of clump 2 and in the region separating clump 8 and 6 on the inward side of the ammonia emission with the region between clump 8 and 6 also being the location of a \HII\ region.\\

\noindent \textbf{W Region:} The region to the west of WR 48a shows a different morphology to previous regions. Here we find 7 isolated clumps of weak emission extending away from the centre of G305 as opposed to large clumps. The strongest \NH\ (1,1) source and the only source of visible \NH\ (2,2) and (3,3) emission are located some distance from the boundary PDR in clump 5. Six \water\ masers were detected, five of which are in close proximity to \NH\ clumps. We note that the PDR in this region appears to be dispersed to the north and south leaving only a small area of PAH emission coincident with clump 14.\\

\subsubsection{{\sc Fellwalker} Clumpfind}

We have chosen to use the STARLINK tool CUPID \footnote{http://starlink.jach.hawaii.edu/starlink/CUPID} to automatically detect clumps in our 2D map, in particular the {\sc Fellwalker} algorithm developed by David Berry \citep{Berry2007}. The {\sc Fellwalker} algorithm works by finding the paths of steepest gradient from each pixel in the image. Starting with the first pixel in the image  each of the surrounding pixels are inspected  to locate the pixel with the highest ascending gradient, this process continues until a peak is located (i.e.~a pixel surrounded by flat or descending gradients). The pixels along this path are assigned an arbitrary integer to represent their connection along a path. All pixels in the image are inspected in a similar process and the image is segmented into clumps by grouping together all paths that lead to the same peak value. The pixels belonging to paths that lead to a single peak are then defined as belonging to that particular clump. For a fuller description of this process see \citet{Berry2007}.
The {\sc Fellwalker} algorithm is best used on images that have had the background subtracted, this has been performed using the background tracing and subtraction algorithm Findback in the CUPID package to subtract the uniform background of 0.06 K from the peak temperature, moment integrated image. The parameters of the detected \NH\ (1,1), (2,2) and (3,3) clumps can be found in Tables 2, 3 and 4. {\sc Fellwaker} is not a widely used clump finding algorithm and so we carried out a number of checks using better known algorithms such as GAUSSCLUMPS and CLUMPFIND to check for consistency, as well as performing manual aperture photometry checks. The {\sc Fellwalker} algorithm gives  very similar results to GAUSSCLUMPS, CLUMPFIND and manual aperture photometry with a significantly higher robustness to varying input parameters \citep[see e.g.][for a discussion of the effect of varying the step size in CLUMPFIND]{Pineda2009}. Due to the simplicity and robustness of {\sc Fellwalker} we chose to use this algorithm to segment our images into clumps and extract the photometry for each clump.

\begin{table*}
\caption{\NH\ (1,1) clump properties, showing central position, velocity, diameter (geometric average), peak main beam temperature and clump averaged main beam temperature. We can only resolve clump diameters of $>2.6$ pc at 4 kpc with the Mopra beam. These values are taken from the spectra averaged over the \NH\ (1,1) clump region defined by {\sc Fellwalker}  }
  \begin{tabular}[h]{ccccccccc}
   \hline Region & Clump No & \multicolumn{2}{c}{Peak}& V$_{\rm LSR}$ & FWHM $\Delta V$ &Diameter&  Peak T$_{\rm MB}$(1,1) & Avg T$_{\rm MB}$(1,1)\\
  & & Galactic (\textit{l}) & Galactic (\textit{b}) & (\kms) & (\kms)  &   (pc) & (K)	& (K) \\
		\hline
	NE	&	1	&	305.23	&	0.29	&	-40.4	&	3.8		&	10.1	&	1.21	&	0.33	\\
		&	3	&	305.19	&  	0.21	&	-42.1	&	4.6	&	5.2	&	0.56	&	0.22	\\	
	\hline
		NW	&	4	&	305.36	&  	0.20	&	-38.8	&	5.6	&	7.6	&	0.55	&	0.22	\\
		&	7	&	305.42  &	0.25	&	-39.8	&	3.1	&	7.4	&	0.46	&	0.19	\\	
		&	9	&	305.54  &	0.34	&	-36.4	&	4.8	&	3.4	&	0.32	&	0.17	\\	
	\hline	
	SE	&	2	&	305.26	&	-0.02	&	-32.3	&	4.9	&	7.3	&	0.74	&	0.28	\\
		&	8	&	305.19  &	0.01	&	-34.0	&	7.1	&	5.2	&	0.43	&	0.15	\\	
		&	6	&	305.14 	& 	0.061	&	-37.4	&	3.4	&	7.0	&	0.45	&	0.21	\\
	\hline
	W	&	5	&	305.82	&  	-0.11	&	-42.7	&	3.0		&	6.0	&	0.60	&	0.22	\\
		&	10	&	305.83 	&	-0.19	&	-34.7	&	3.5		&	3.2	&	0.28	&	0.12	\\
		&	11	&	305.89 	&	0.02	&	-35.1	&	3.2		&	3.9	&	0.27	&	0.16	\\
		&	12	&	305.69  &	0.04	&	-37.4	&	1.7		&	2.9	&	0.34 &	0.18	\\
		&	13	&	305.77  &	-0.25	&	-32.3	&	4.1		&	3.9	&	0.34 &	0.14	\\
		&	14	&	305.56  &	0.02	&	-40.4	&	3.9		&	3.8	&	0.34	&	0.16	\\
		&	15	&	305.48 	&	-0.10	&	-39.3	&	2.6		&	2.9	&	0.31 &	0.19	\\
	
\hline
\end{tabular}
\end{table*}

\begin{table*}
\caption{\NH\ (2,2) clump properties, showing central position, velocity, diameter (geometric average), peak main beam temperature and clump averaged main beam temperature, these values are taken from the spectra averaged over the \NH\ (1,1) clump region defined by {\sc Fellwalker} .}
  \begin{tabular}[h]{cccccccccc}
  
   \hline Region & Clump No & \multicolumn{2}{c}{Peak} &V$_{\rm LSR}$ &FWHM $\Delta V$&Diameter&  Peak T$_{\rm MB}$(2,2) & Avg T$_{\rm MB}$(2,2)\\
  & & Galactic (\textit{l}) & Galactic (\textit{b})& (\kms) &(\kms)& (pc) & (K)	& (K) \\
 	\hline
	NE	&	1	&	305.23 & 	0.28	&	-39.9	&	3.8	&	6.3	&	0.66 	&	0.17	\\
		&	3	&	305.20 &	0.21	&	-41.6	&	6.2	&	4.3	&	0.32	&	0.14	\\
		\hline
	NW	&	4	&	305.36 & 	0.19	&	-38.6	&	6.2	&	4.5	&	0.34	&	0.14	\\
		&	7	&	305.42 &	0.25	&	-39.2	&	2.9	&	$<2.6$	&	0.39	&	0.09	\\
		&	9	&	305.49	&	0.35	&	-36.3	&	6.3	&	$<2.6$	&	0.30	&	 0.07	\\
	\hline
	SE	&	2	&	305.26 &	-0.02	&	-32.1	&	5.0	&	5.2 &	0.46	&	0.14	\\
		&	8	&	305.19 &	0.00	&	-33.5	&	6.3	&	3.0	&	0.31	&	0.10	\\
		&	6	&	305.15 &	0.03	&	-37.1	&	3.7	&	$<2.6$	&	0.45	&	0.10	\\
	\hline
	W	&	5	&	305.84 &	-0.07	&	-41.7	&	2.5	&	2.6	&	0.33	&	0.08	\\
		&	12	&	305.69 &	0.06	&	-37.7	&	3.2	&	$<2.6$	&	0.23		& 0.07\\
		&	13	&	305.89	&	-0.25	&	-32.3	&	5.1	&	$<2.6$	&	0.25	&	0.07	\\
		&	14	&	305.57	&	0.028	&	-39.2	&	4.4	&	$<2.6$	&	0.26	&	0.08	\\

\hline
\end{tabular}
\end{table*}

\begin{table*}
\caption{\NH\ (3,3) clump properties, showing central position, velocity, diameter (geometric average), peak main beam temperature and clump averaged main beam temperature. Clump values are derived from the \NH\ (3,3) source averaged spectra Gaussian fit}
  \begin{tabular}[h]{cccccccccc}
  
   \hline Region & Clump No & \multicolumn{2}{c}{Peak} &V$_{\rm LSR}$ &FWHM $\Delta V$&Diameter&  Peak T$_{\rm MB}$(3,3) & Avg T$_{\rm MB}$(3,3)\\
  & & Galactic (\textit{l}) & Galactic (\textit{b})& (\kms) &(\kms)& (pc) & (K)	& (K) \\
 	\hline
	NE	&	1	&	305.22 & 	0.28	&	-40.2	&	4.8	&	5.7	&	0.42	&	0.20	\\
		&	3	&	305.20 &	0.21	&	-42.0	&	7.6	&	3.8	&	0.36	&	0.20	\\
		\hline
	NW	&	4	&	305.35 & 	0.20	&	-38.6	&	7.1	&	4.9	&	0.36	&	0.20	\\
	\hline
	SE	&	2	&	305.25 &	-0.02	&	-32.2	&	5.9	&	4.4 &	0.42	&	0.15	\\
		&	8	&	305.19 &	0.00	&	-33.6	&	10.0&	3.4	&	0.23	&	0.08	\\
	\hline
	W	&	5	&	305.81 &	-0.1	&	-41.9	&	1.4	&	$<2.6$	&	0.28	&	0.14	\\

\hline
\end{tabular}
\end{table*}

\subsection{\NH\ clump spectra}

In Fig. 3, we present source averaged \NH\ (1,1) and (2,2) spectra for each clump region defined by {\sc Fellwalker} (shown as a grey outline surrounding the contours in Fig.2). In Fig. 4, we present the \NH\ (3,3) spectra averaged over the (3,3) emission region. All spectra have been Hanning smoothed to improve the noise to $\sim0.01$ K per $\sim0.8$ \kms channel. We have applied the same \NH\ (1,1) clump area results to the \NH\ (2,2) emission, therefore the (1,1) and (2,2) clump spectra are averaged over the same area, this allows us to make comparisons between the transitions. We should only expect to see seven strong detections in the \NH\ (2,2) and even fewer in the (3,3) transition that are coincident with \NH\ (1,1) emission, as seen in the peak temperature contour map Fig. 2. This is not the case, Fig. 3 clearly shows faint \NH\ (2,2) emission in twelve clumps and we see 6 \NH\ (3,3) detections in Fig. 4, this is due to the improved signal to noise from applying Hanning smoothing. We note the velocity of the \NH\ (2,2) and (3,3) emission appears at approximately the same velocity as the \NH\ (1,1) emission in all cases, for the particularly weak \NH\ (2,2) and (3,3) emission this provide supporting evidence that it is indeed real. The \NH\ (3,3) emission area is much smaller in  geometric size than the (1,1) emission, therefore averaging the (3,3) spectra over the (1,1) area significantly lowers the temperature of the (3,3) emission due to beam dilution. We  make use of \NH\ (3,3) emission to highlight  areas of higher temperature and for this reason we have created \NH\ (3,3) source averaged spectra shown in Fig. 4 in order to better highlight the warmer clumps.

The hyperfine nature of the \NH\ (1,1) inversion transition provides a good indicator of optical depth through the ratio of intensity between the main and satellite lines. For this reason we have applied hyperfine fitting to all the \NH\ (1,1) spectra. Due to insufficient signal to noise, we are unable to resolve hyperfine structure in the \NH\ (2,2) emission and so we have applied Gaussian fitting. We only fit a Gaussian to the \NH\ (2,2) and (3,3) data if we can be confident that it is a true detection and the peak emission is $>3\sigma$, thus only clumps 10, 11 and 15 are left with no (2,2).

Profiles were fitted using the IDL functions CURVEFIT for \NH\ (1,1) and GAUSSFIT for the \NH\ (2,2) and (3,3) lines. The model line profiles are over plotted in red in Fig. 3 and 4. The GAUSSFIT function computes a non-linear least-squares fit to the data. The CURVEFIT function uses a gradient-expansion algorithm to compute a non-linear least squares fit to a user-supplied function with an arbitrary number of parameters. Iterations are performed until the chi square changes by a specified amount, or until a maximum number of iterations have been performed. To fit the ammonia profiles we assume all the components have equal excitation temperatures and that the ammonia line widths and separations are identical to laboratory values. 

We find line widths range from 1.7 \kms\ to 10.0 \kms. The thermal part of the line width accounts for only 0.28 \kms\  (for T$_{\rm kin}=30$ K) we therefore assume the line widths are largely dominated by non-thermal turbulent motion. It can be clearly seen that several spectra exhibit blended main line and first satellite lines, this may be caused by emission from several unresolved clumps with differing velocities. The \NH\ (3,3) line widths are found to be wider than the lower transition counterparts for all clumps with the exception of clump 5.

\begin{figure*}
\includegraphics[width=0.26\textwidth]{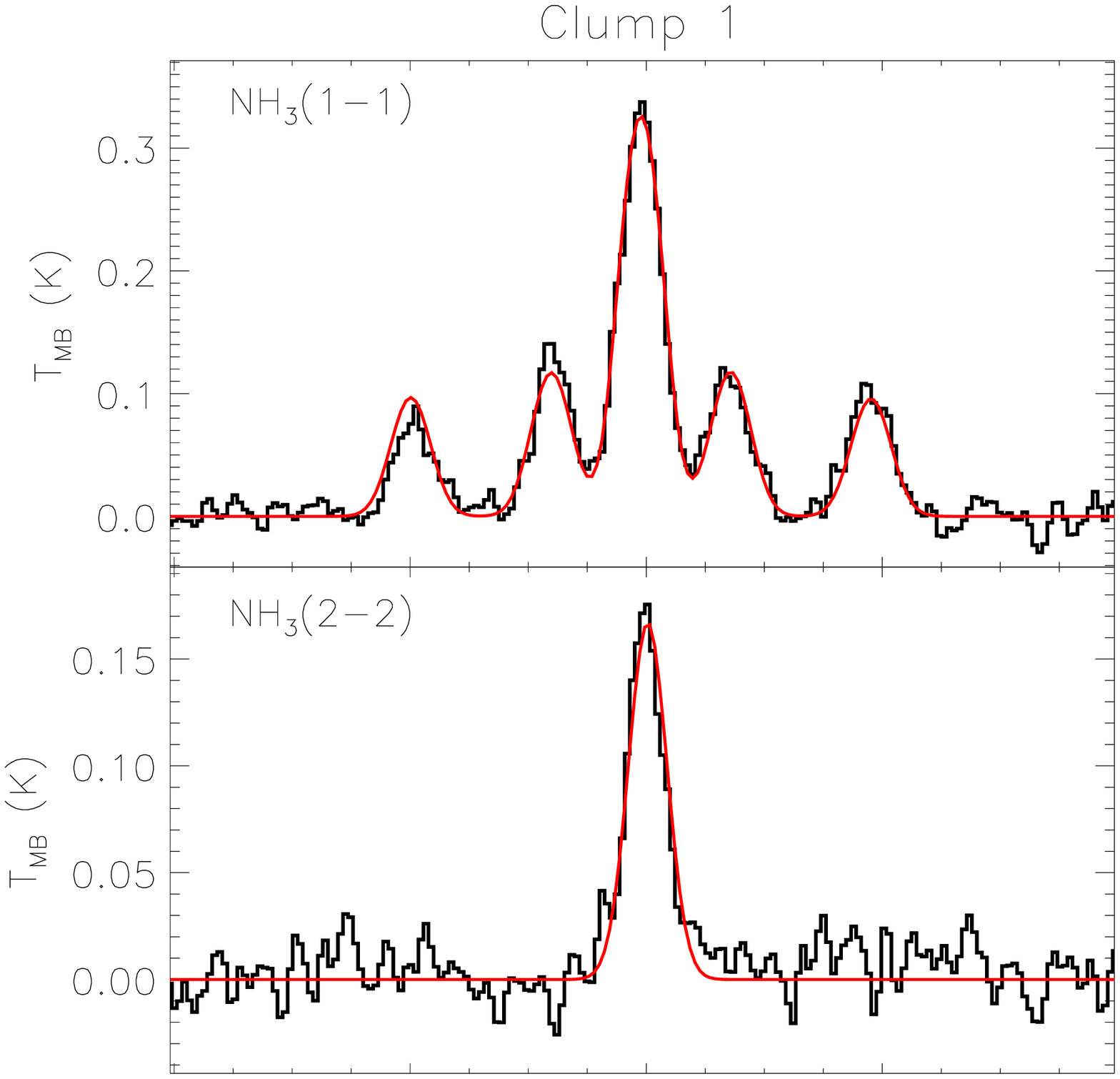} 
\includegraphics[width=0.26\textwidth]{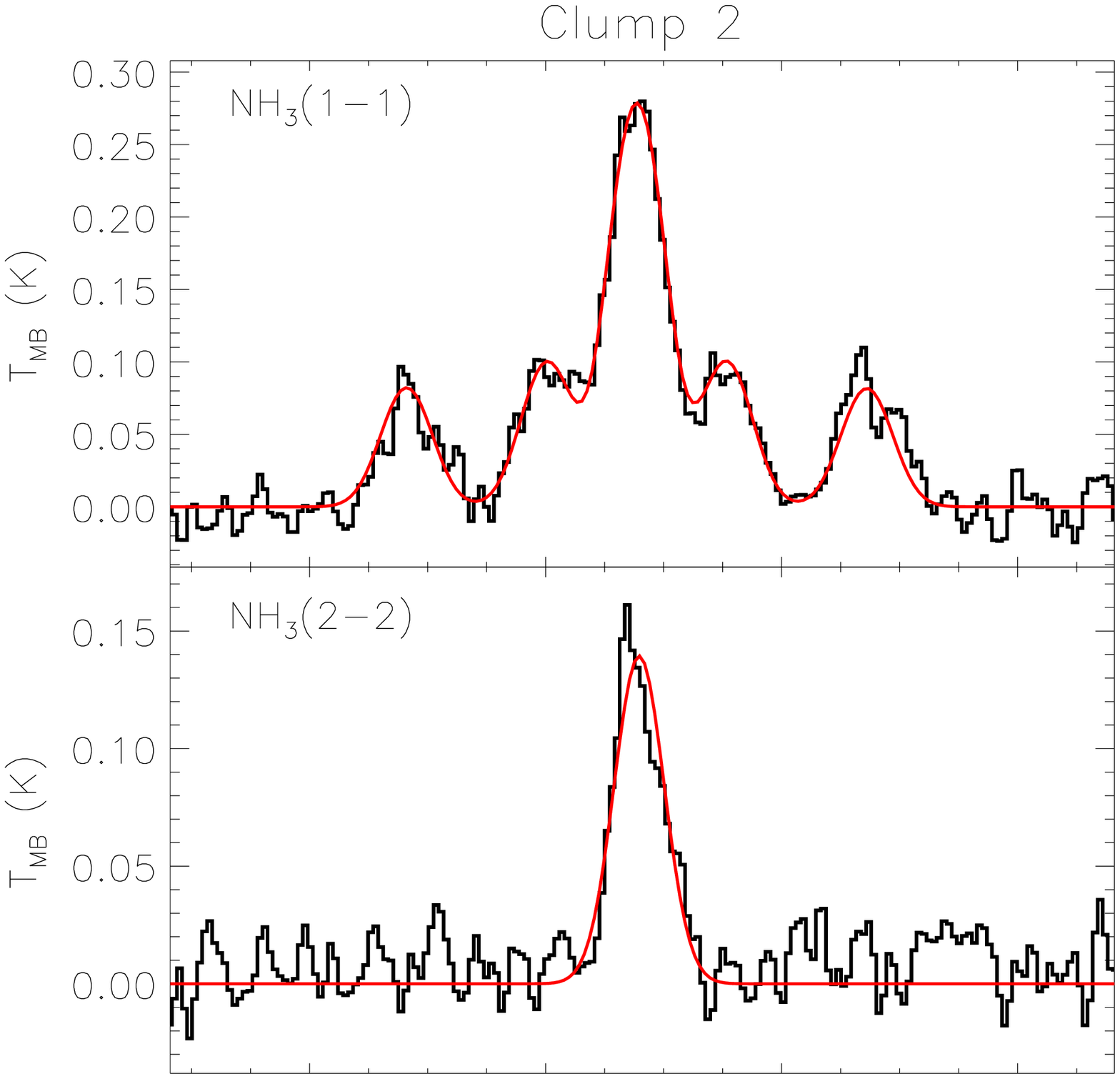} 
\includegraphics[width=0.26\textwidth]{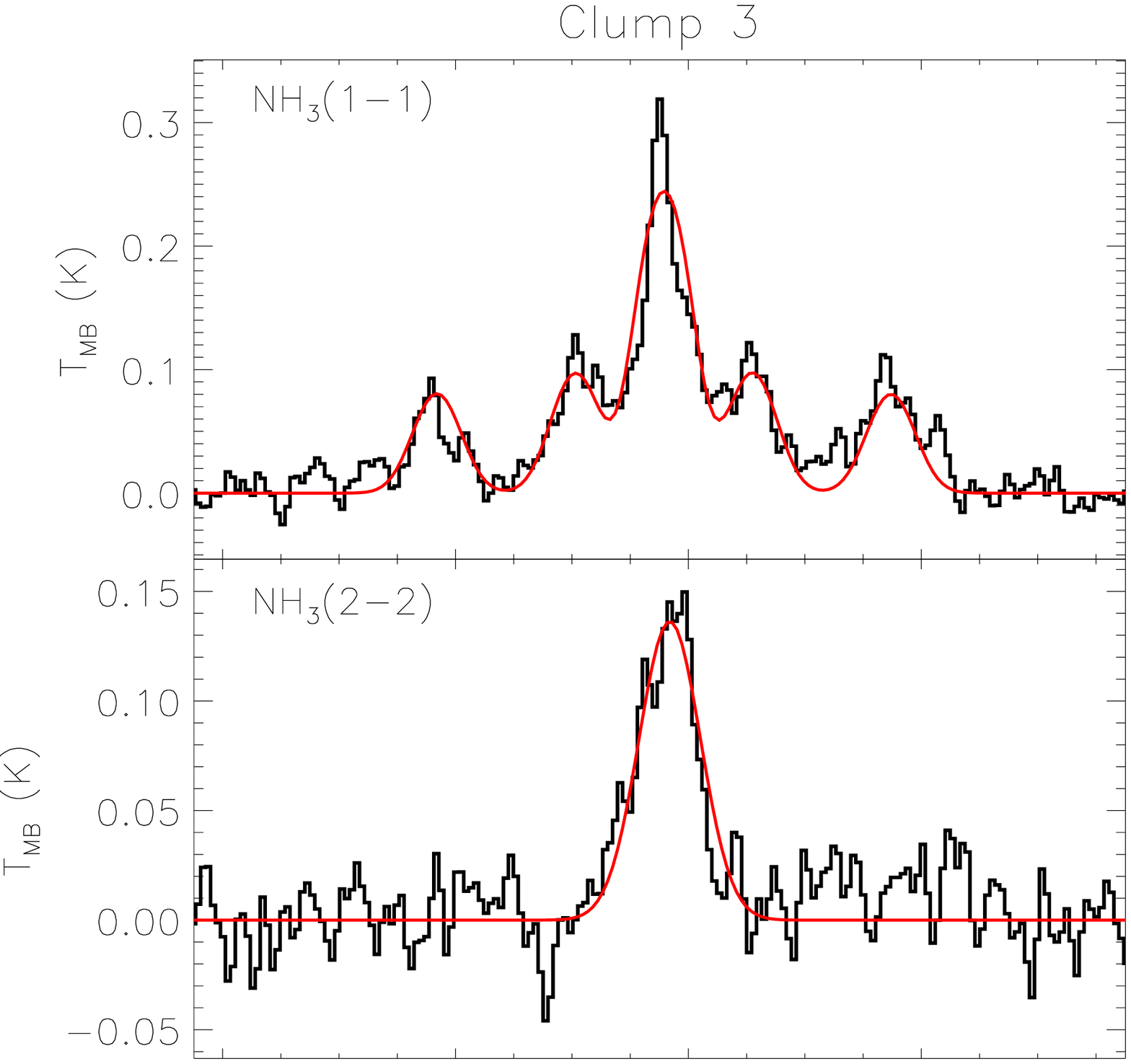} 
\includegraphics[width=0.26\textwidth]{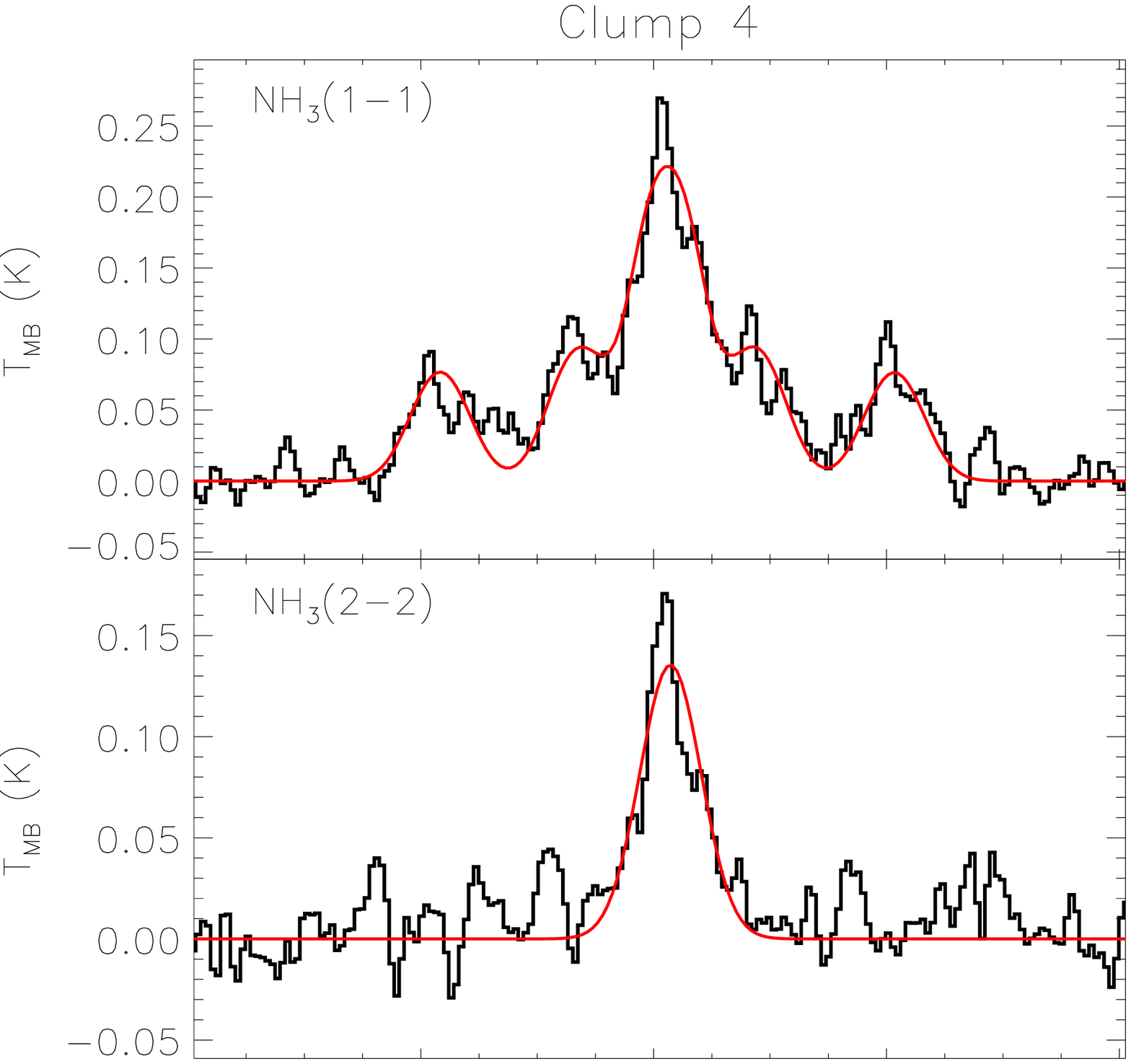} 
\includegraphics[width=0.26\textwidth]{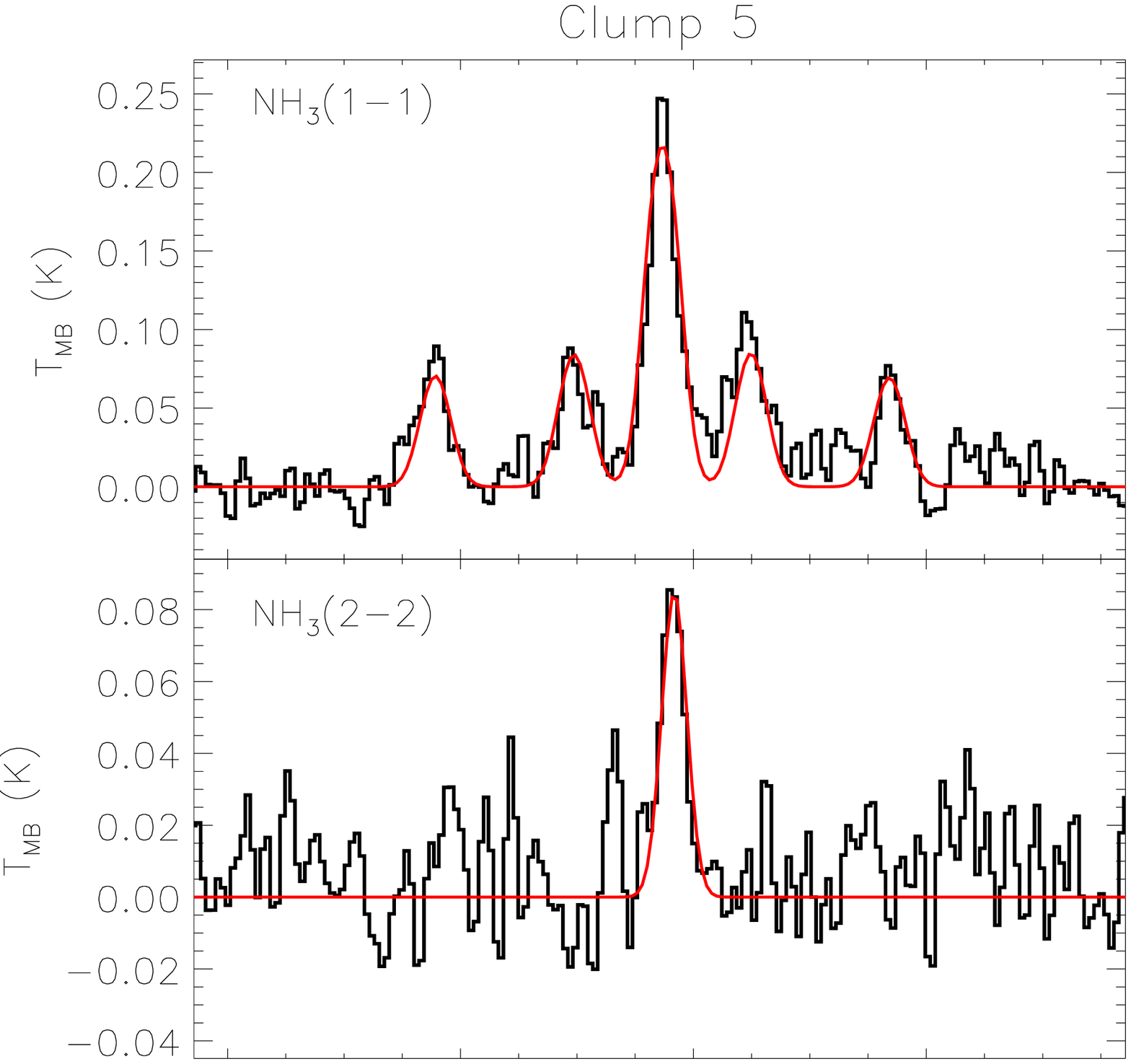} 
\includegraphics[width=0.26\textwidth]{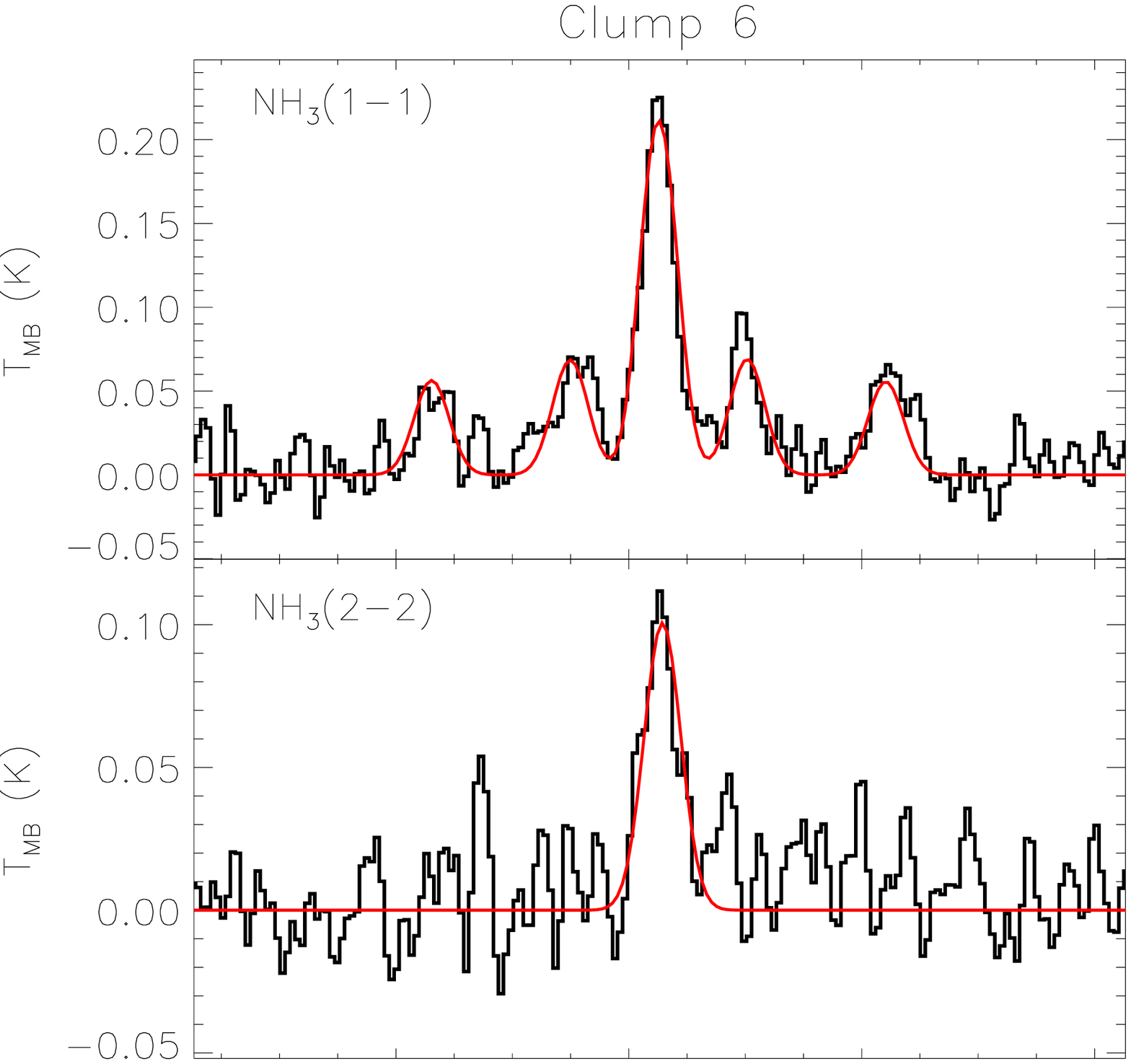}
\includegraphics[width=0.26\textwidth]{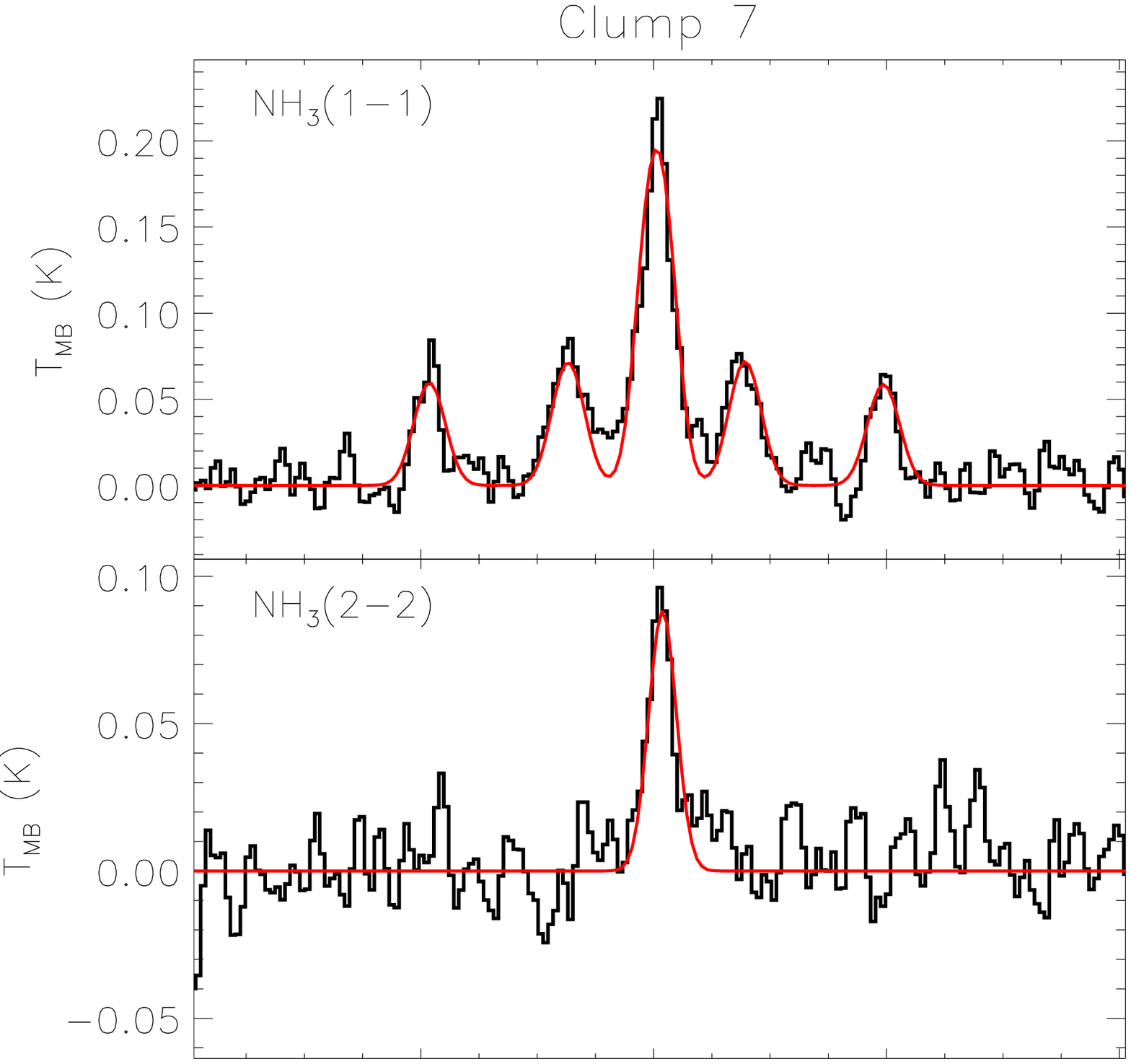}
\includegraphics[width=0.26\textwidth]{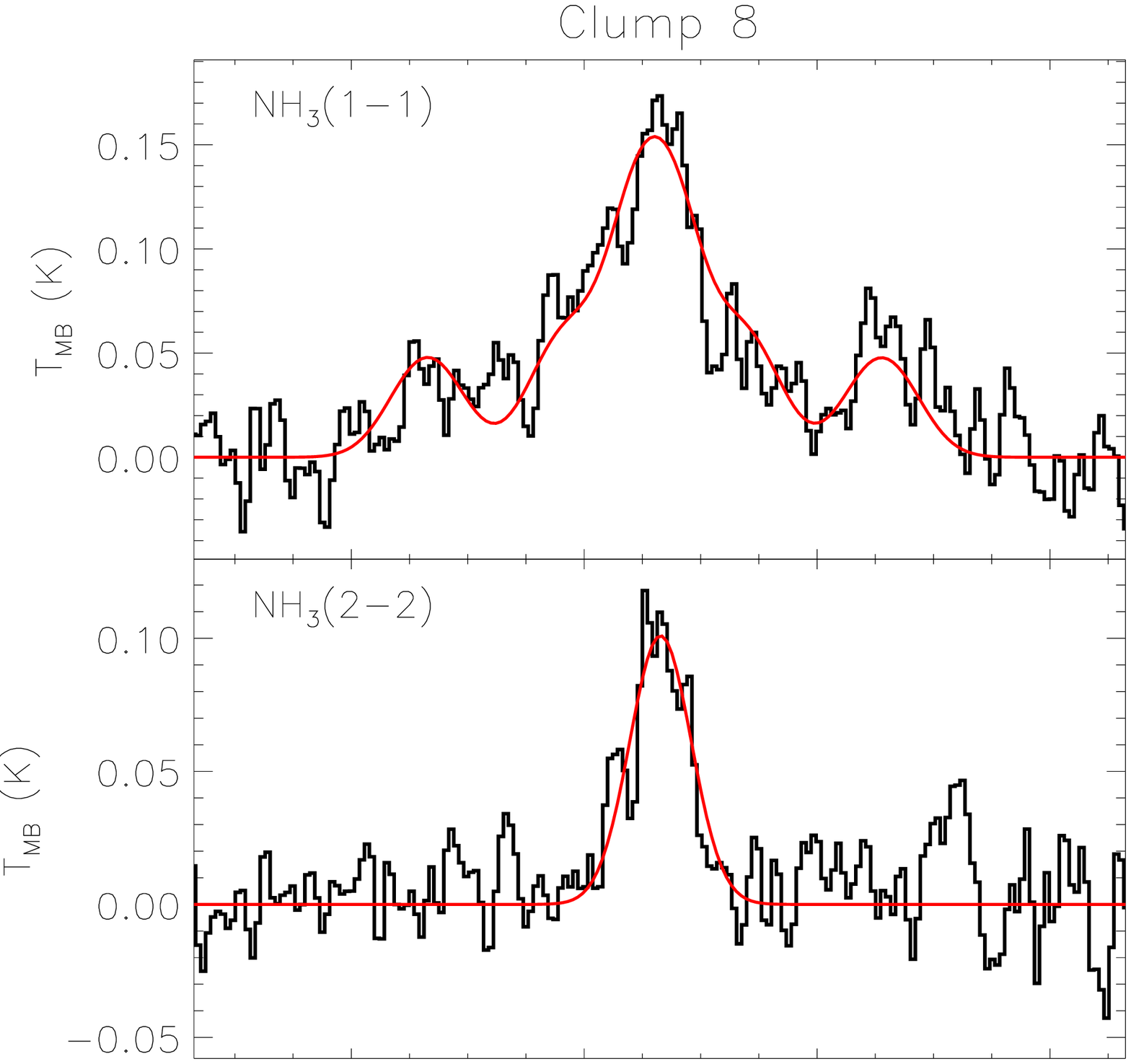}
\includegraphics[width=0.26\textwidth]{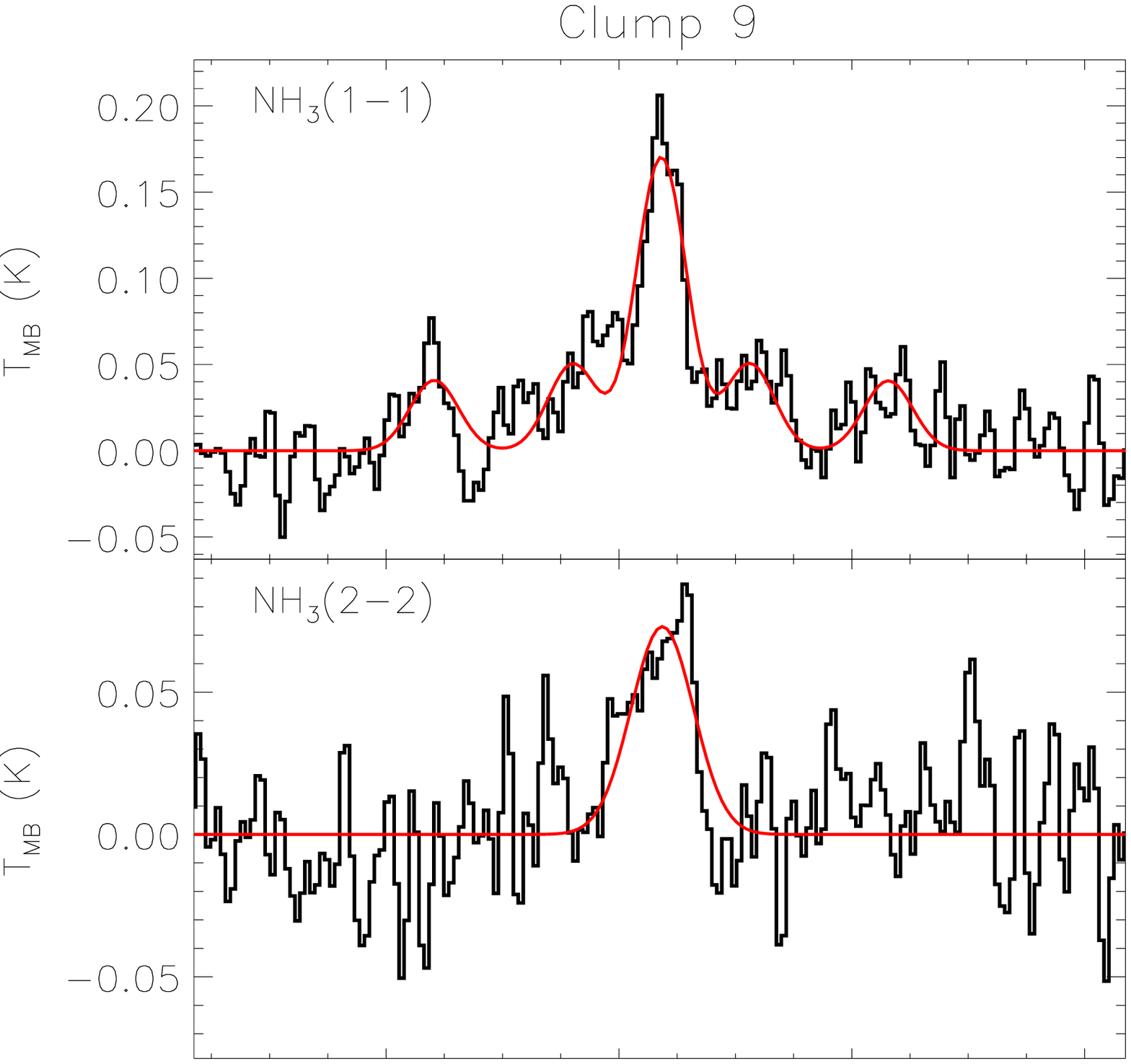}
\includegraphics[width=0.26\textwidth]{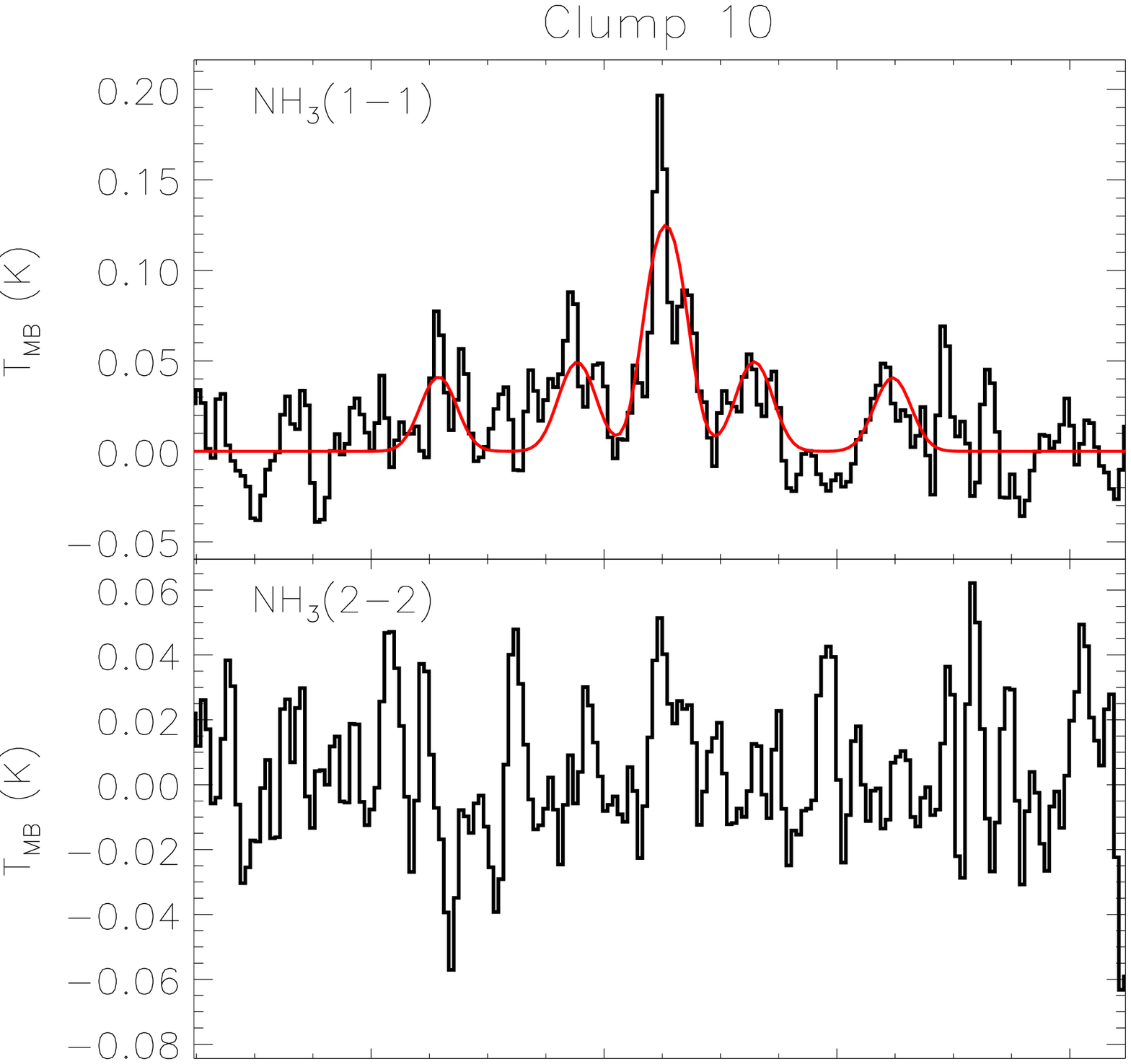}
\includegraphics[width=0.26\textwidth]{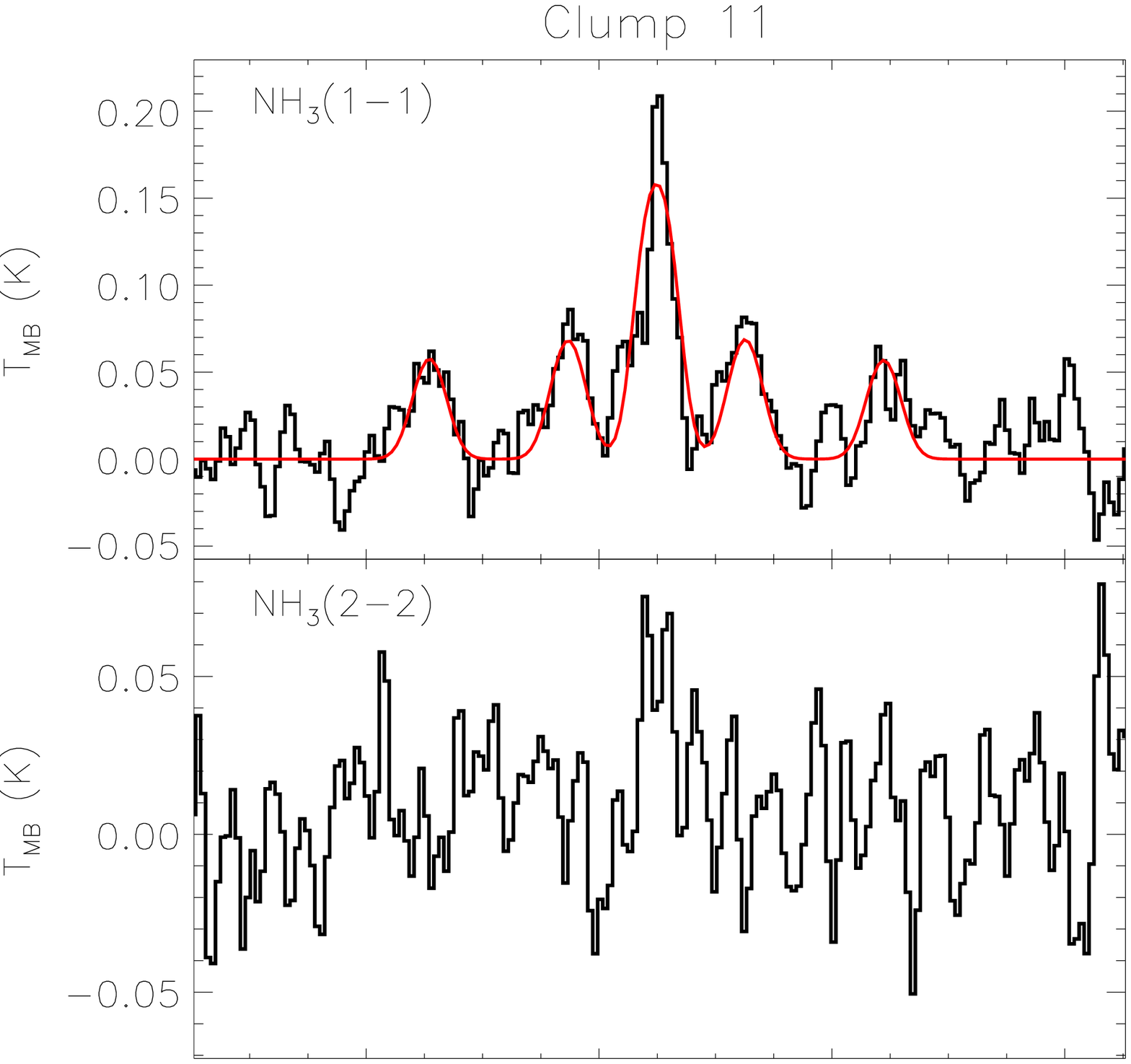}
\includegraphics[width=0.26\textwidth]{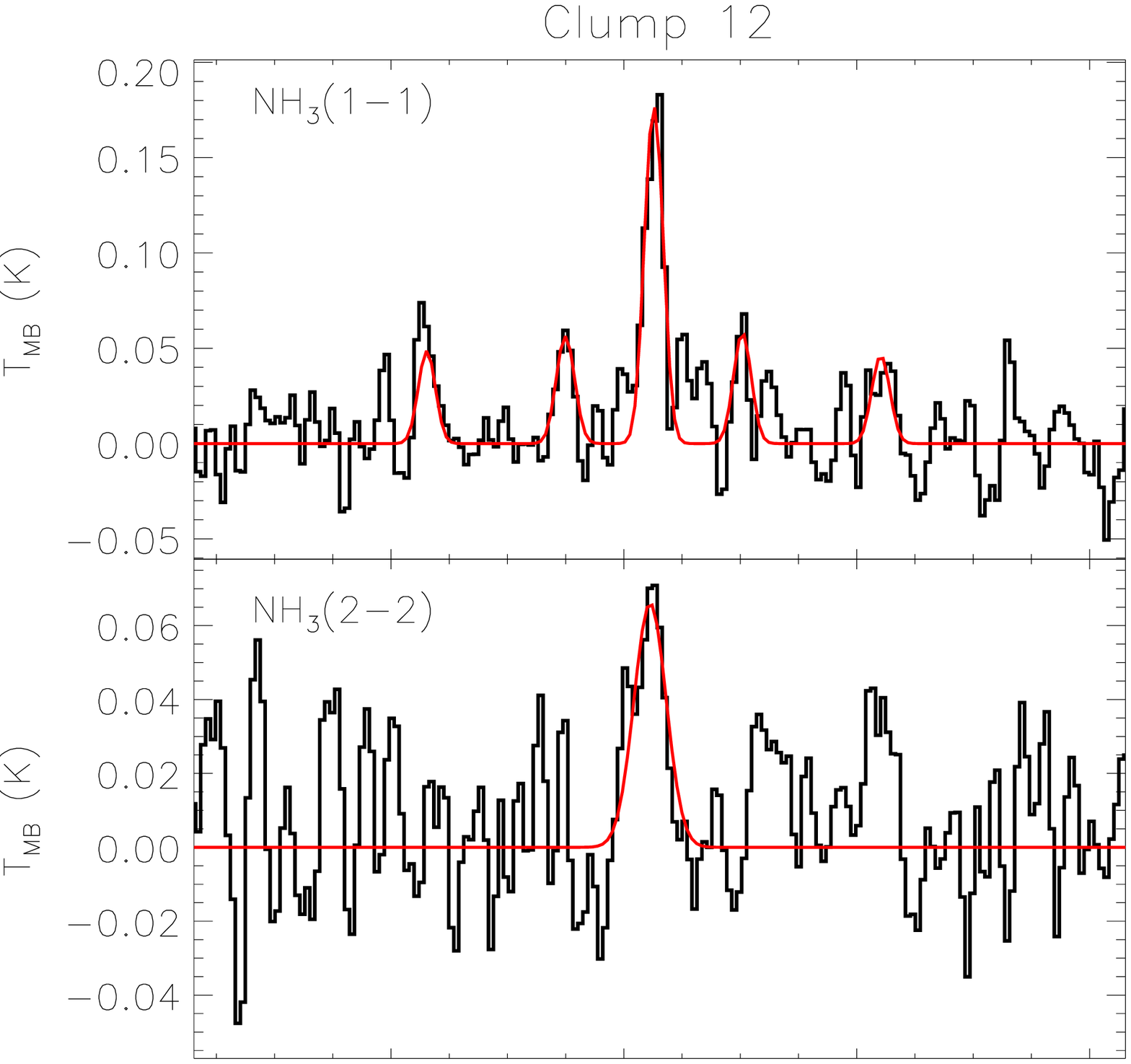}
\includegraphics[trim = 0 15 0 0 , width=0.26\textwidth]{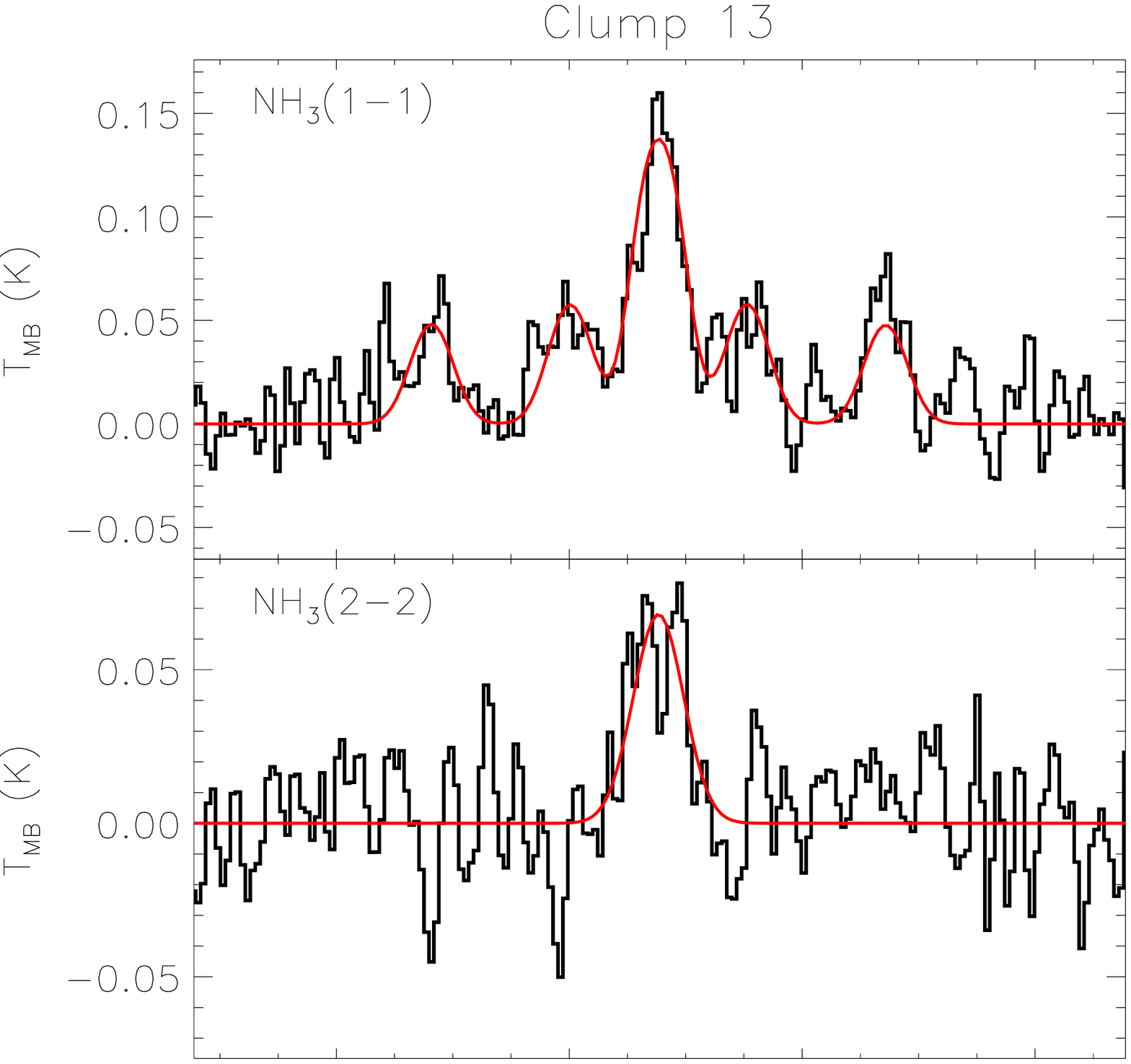}
\includegraphics[trim = 0 15 0 0 ,width=0.26\textwidth]{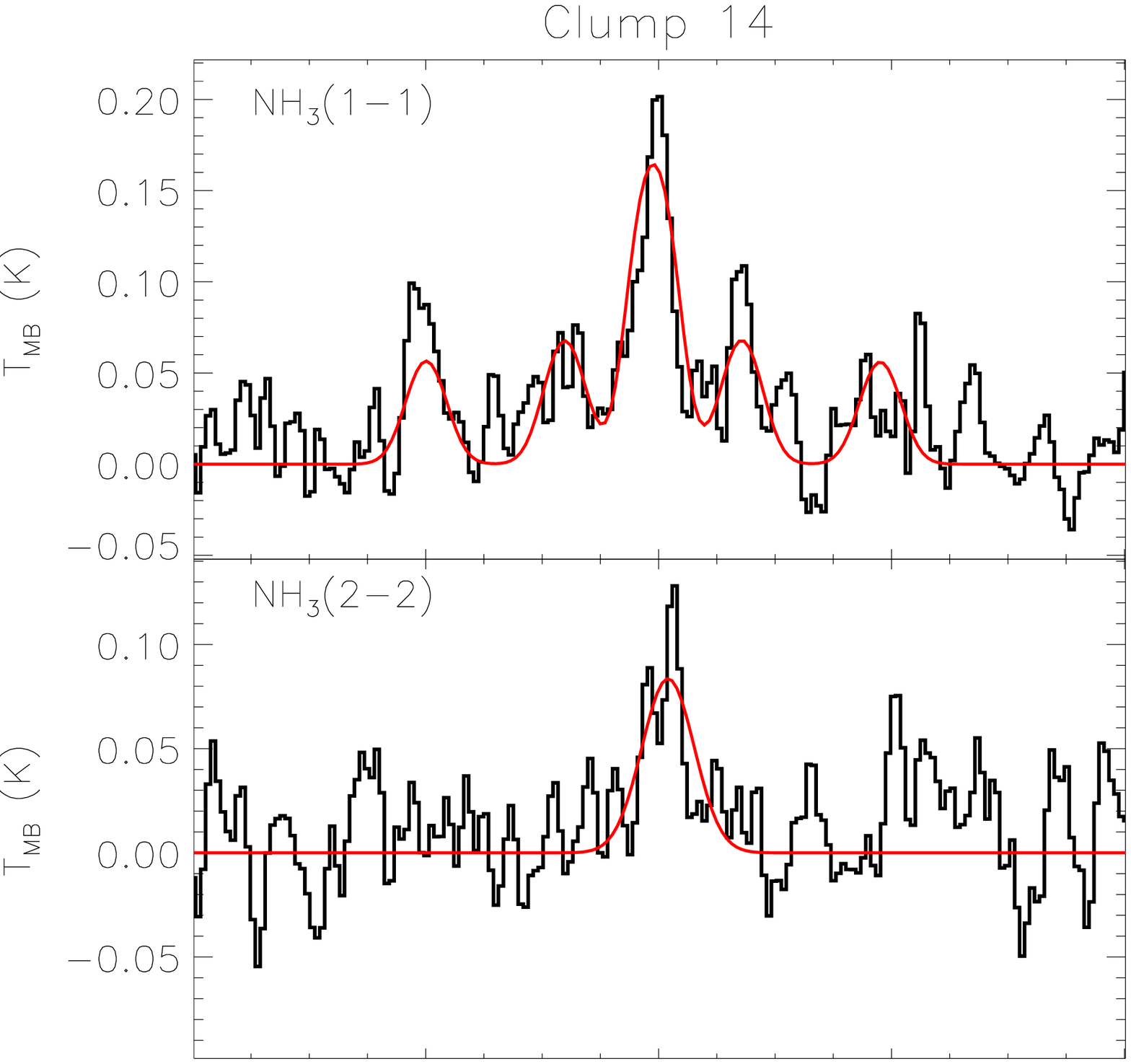}
\includegraphics[trim = 0 15 0 0 ,width=0.26\textwidth]{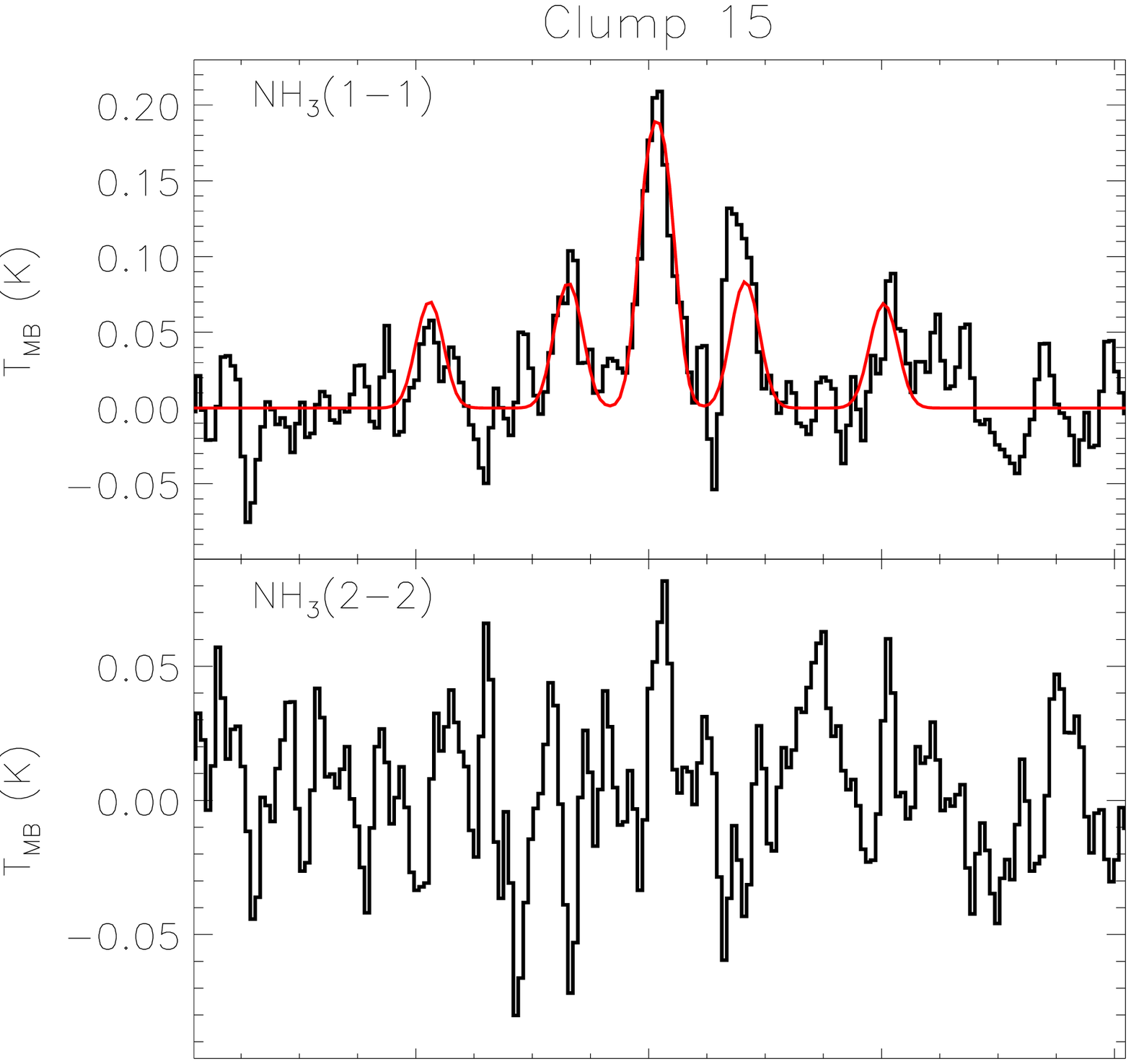}
\caption{Clump averaged spectra of the 15 \NH\ clumps towards G305. Clumps have been integrated spatially over the clump area defined by {\sc Fellwalker} and Hanning smoothed to provide a sensitivity of $\sim0.01$ K per $\sim0.8$ \kms channel. Hyperfine fitting is applied to the (1,1) emission and Gaussian fitting to the (2,2), shown as red overplots.}
\label{Spectra of the 15 $NH_3$ clumps identified by Fellwalker in section 3.1.1}
\end{figure*}
\begin{figure*}
\includegraphics[width=0.26\textwidth]{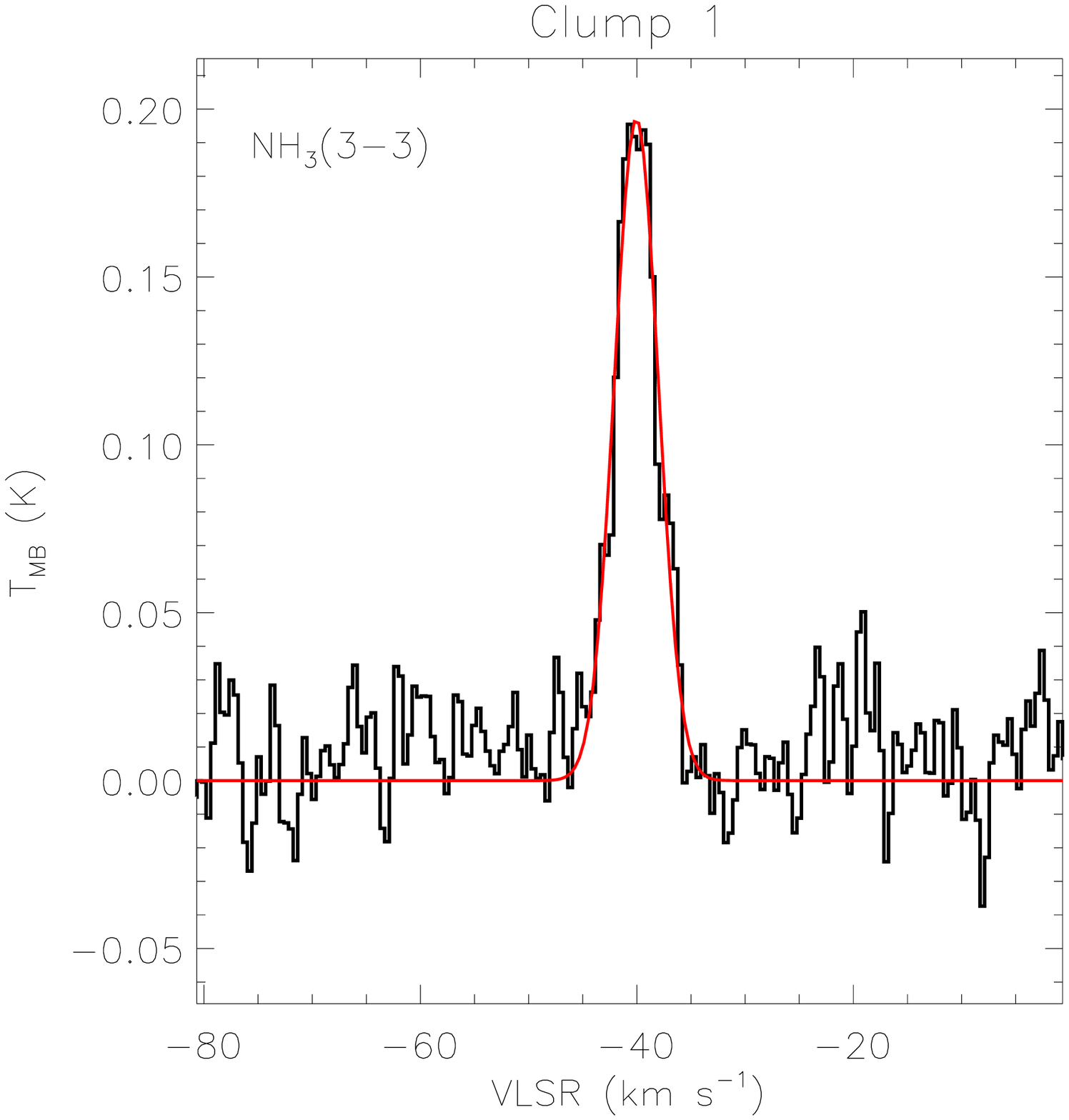} 
\includegraphics[width=0.26\textwidth]{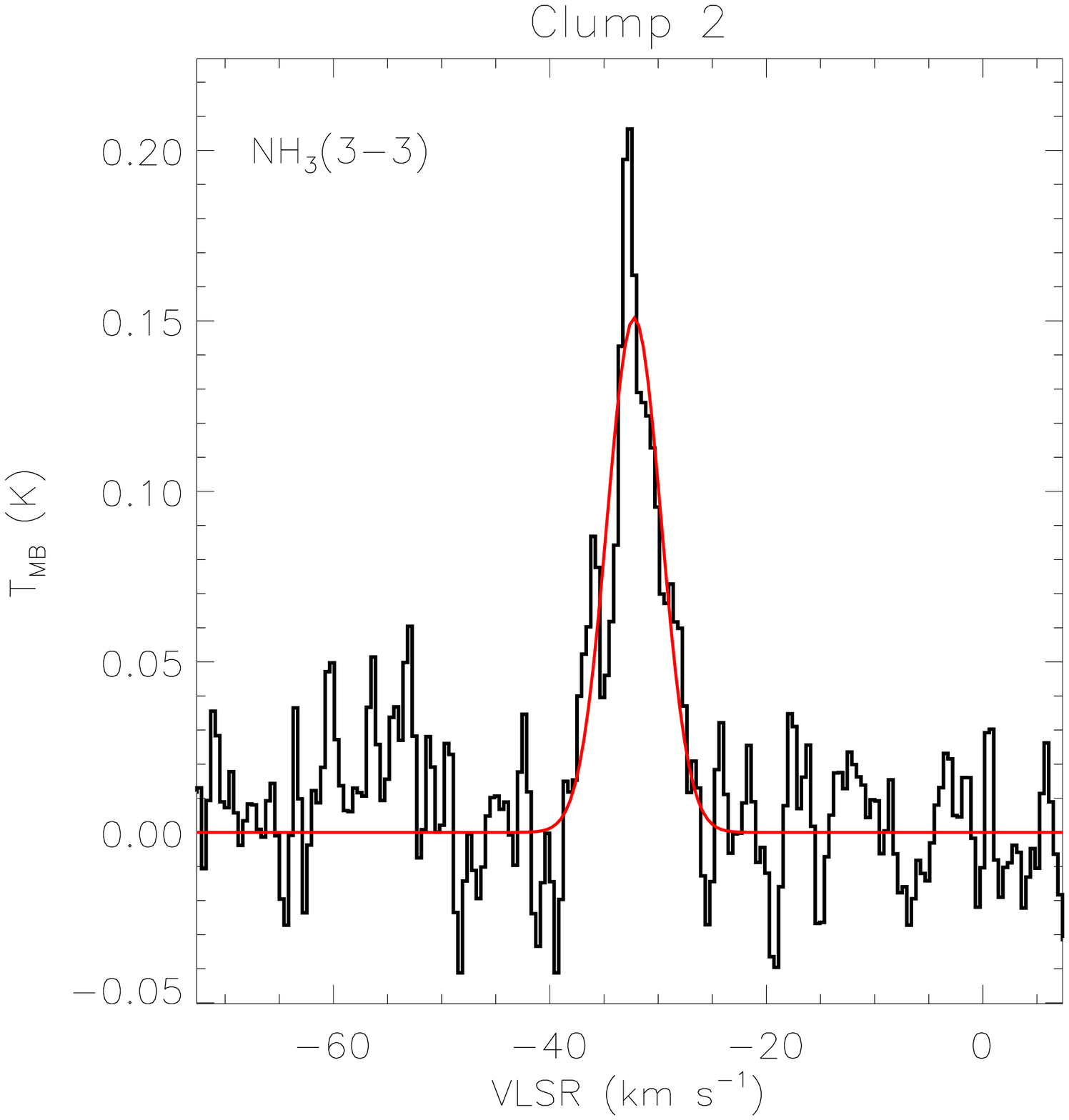} 
\includegraphics[width=0.26\textwidth]{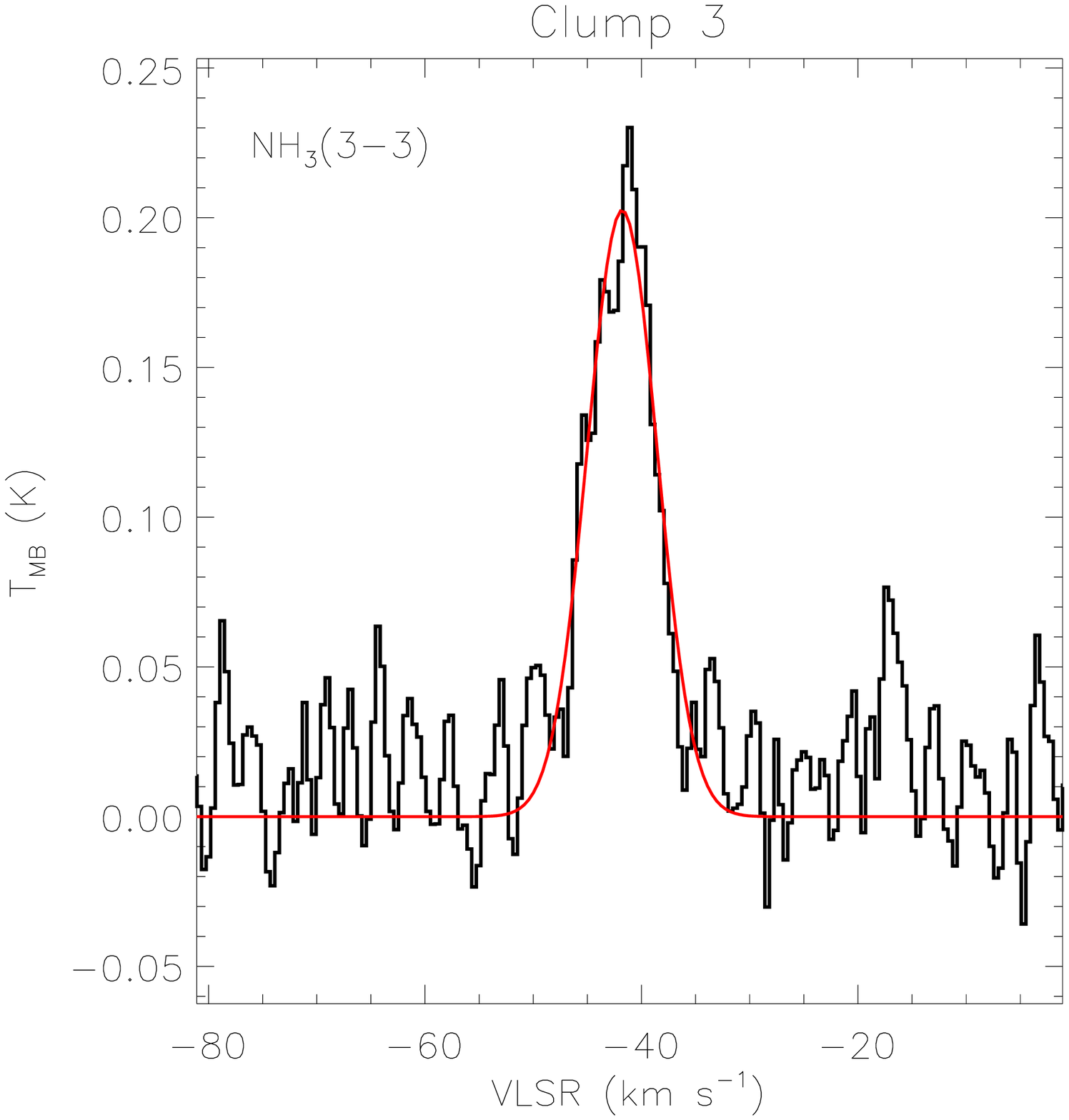} 
\includegraphics[width=0.26\textwidth]{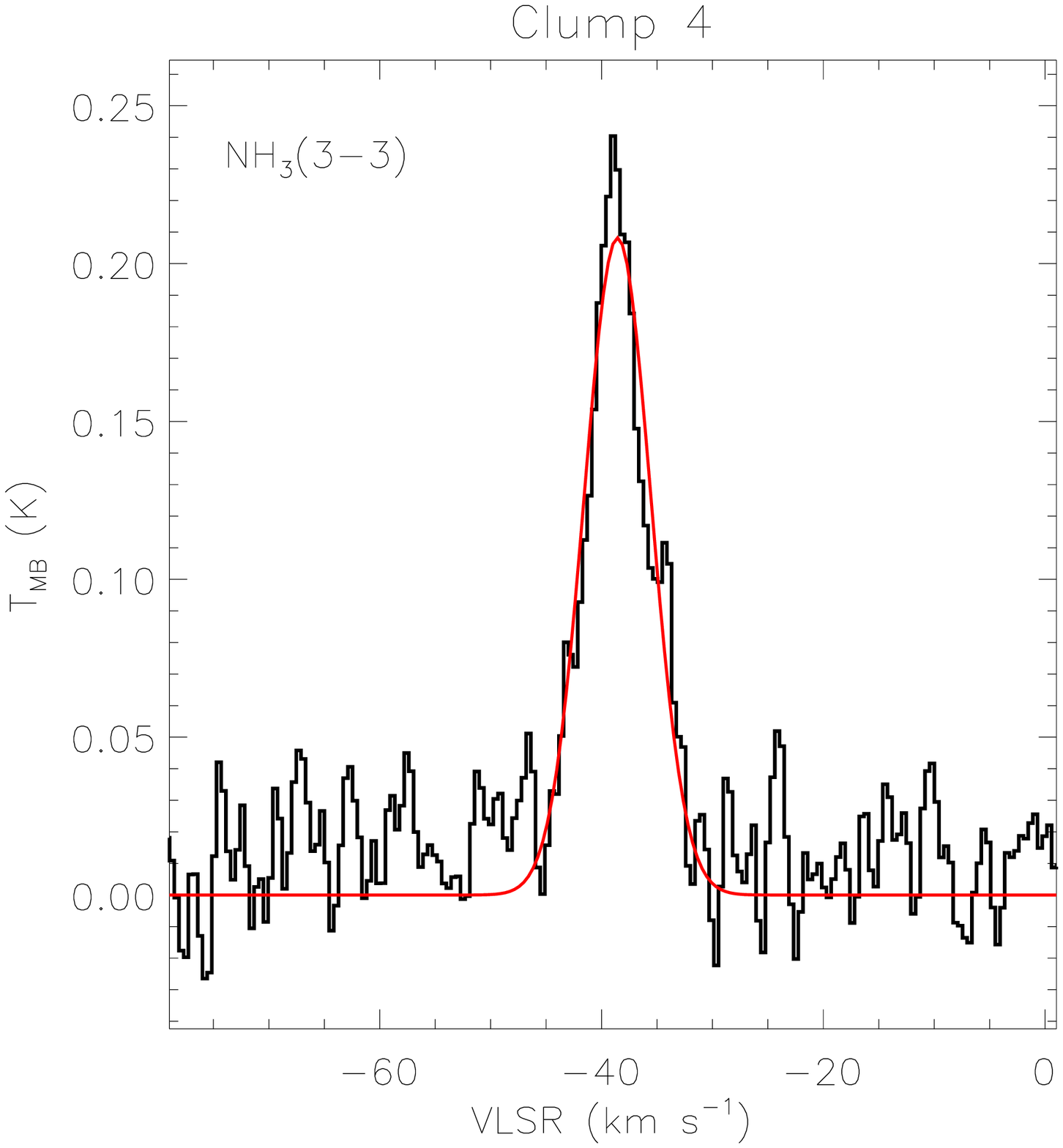} 
\includegraphics[width=0.26\textwidth]{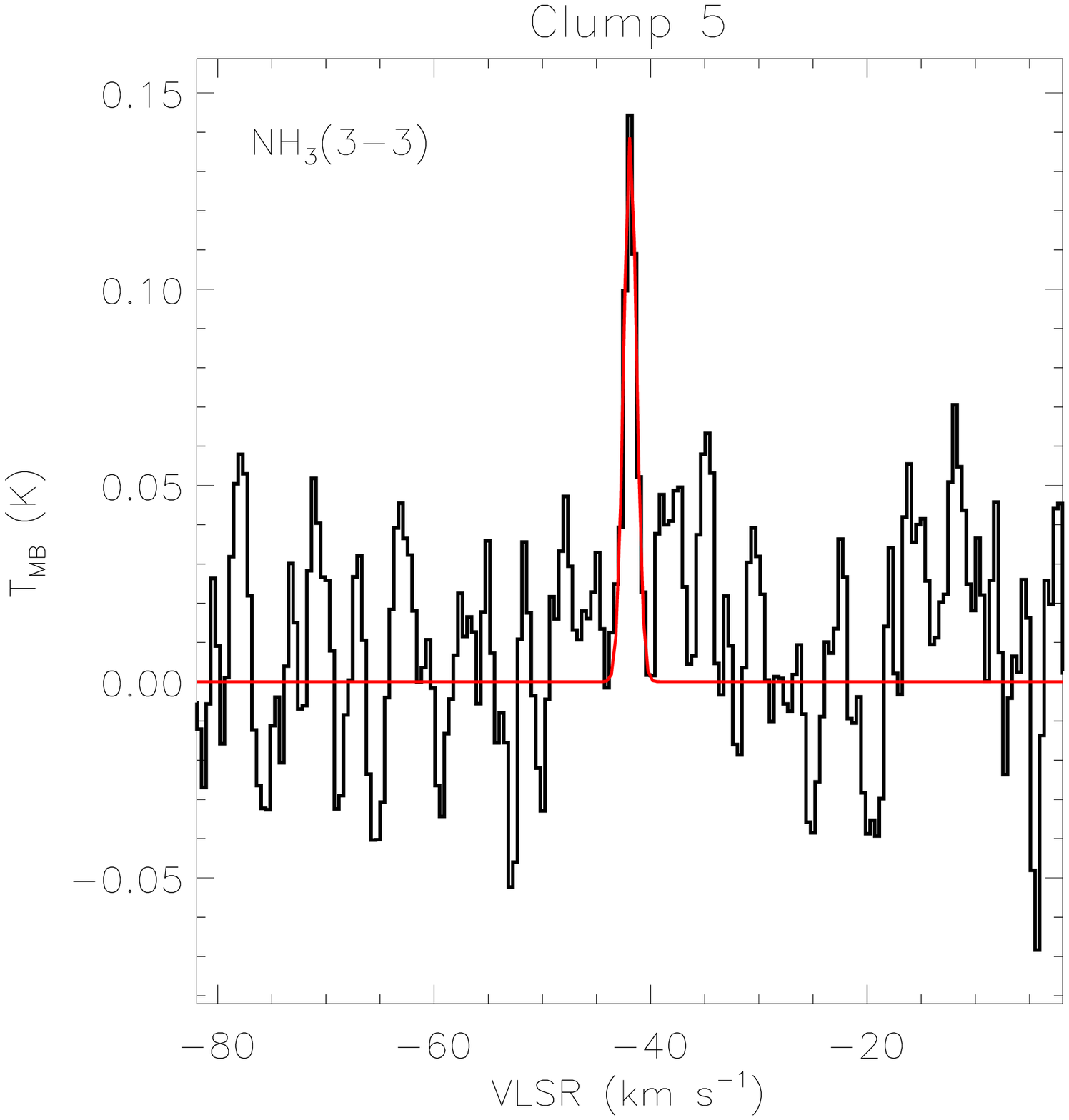} 
\includegraphics[width=0.26\textwidth]{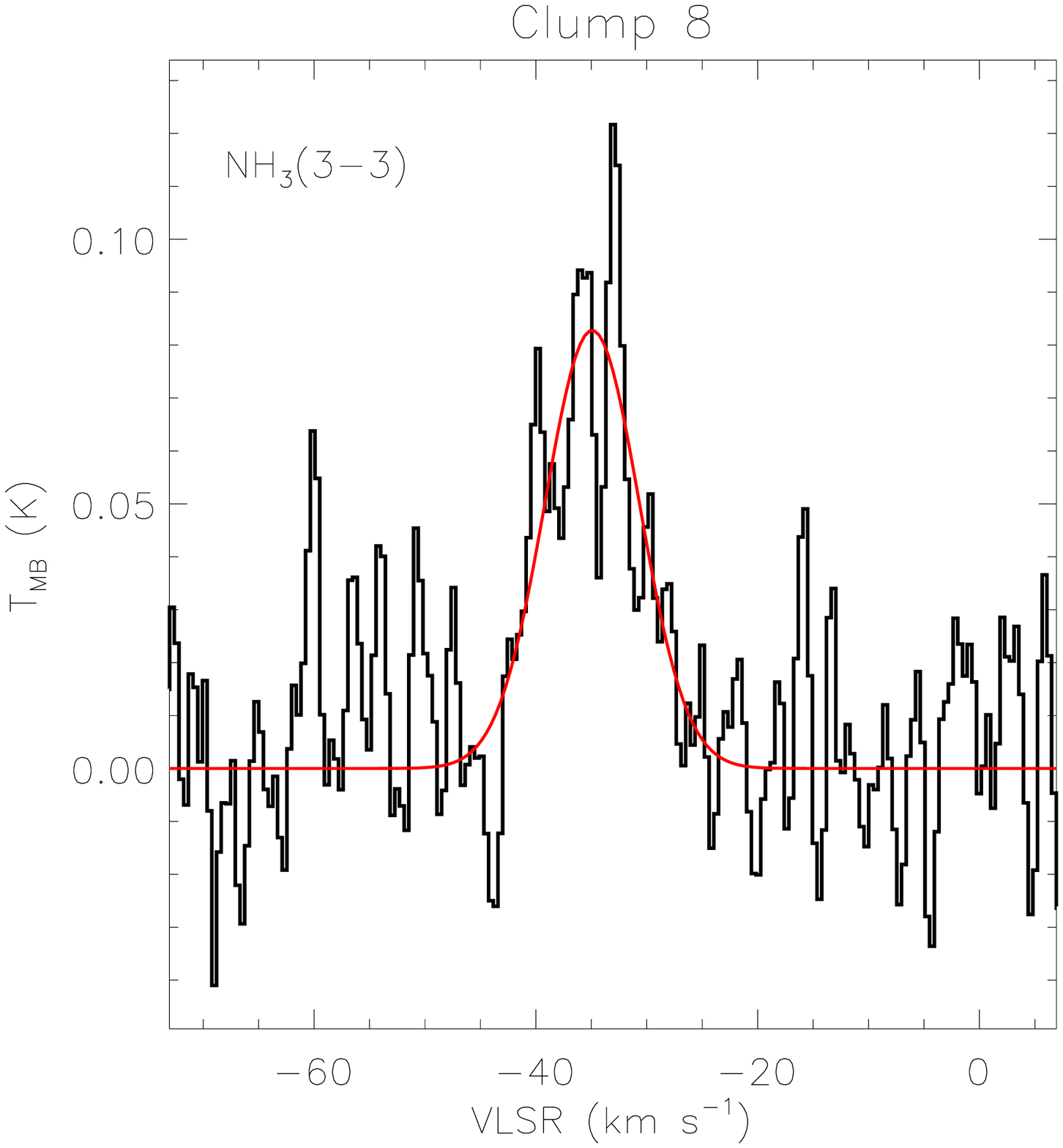}
\caption{Source averaged spectra of the 6 \NH\ (3,3) clumps towards G305. Clumps have been integrated spatially over the \NH\ (3,3) emission area, spectra have been Hanning smoothed to provide a sensitivity of $\sim0.01$ K per $\sim0.8$ \kms channel. Gaussian fitting is applied, shown as red overplots.}
\label{Spectra of the 6 $NH_3$ (3,3) clumps}
\end{figure*}

\subsection{\water\ masers}

Sixteen \water\ masers were detected towards G305, their positions, velocity (LSR), velocity range and peak flux can be found in Table 5. It is important to note that due to the large size of the Mopra beam ($\sim2$\arcmin), we are currently unable to determine the location of the emission to better than a few 10s of arcseconds. However follow up observations currently being obtained by the HOPS survey \citep{Walsh2008} will accurately determine the locations to within a few arc-seconds. We make use of a recent study into the efficiency of Mopra by Urquhart et al (submitted 2010), specifically Table 3, to convert \water\ maser antenna temperatures to Jy using the conversion factor of 12.4. We find 6 new \water\ masers in the region, \cite{Walsh2008} show a detection of two \water\ maser at $l=305.55, b=-0.02$ and $l=305.85,b=+0.07$ which we did not detect. The only apparent explanation for this is \water\ maser variability, since our observational setup mimicked the HOPS survey and we obtained deeper, less noisy maps we should expect to detect all of the \water\ masers reported in \cite{Walsh2008} and more. Future HOPS observations should clarify this ambiguity.
\water\ maser emission was detected across a broad velocity range between -120.8 and -10.4 \kms\ with a peak intensity varying between 2.4 and 1087.8  Jy, these values are characteristic of \water\ maser emission \citep{Breen2007}. The peak V$_{\rm LSR}$ and velocity range match well with the velocity of the \NH\ emission, it is clear that all detected \water\ masers are associated with G305 with the exception of maser 12, which has a peak velocity of -97.6 \kms and a velocity range that does not coincide with the G305 complex.

\begin{table*}
\caption{\water\ maser positions, peak velocity, velocity range and peak flux. Asterisks above the maser number show new detections over those identified by Walsh et al (2008). Previously detected \water\ masers are detailed in Table 2 in Walsh et al (2008)}
  \begin{tabular}[h]{cccccccc}
   \hline Region & Maser & \multicolumn{2}{c}{Peak} & V$_{\rm LSR}$ & Velocity & Peak \\
   &	Number &Galactic &Galactic		&Peak	& Range & Flux \\
   		 & &(\textit{l})& (\textit{b}) &(\kms) &(\kms)& (Jy) \\
 	\hline
NE	&	$1^*$	&	305.22  &	0.28   	&	-34.0	&	-37.7, -33.1	&	8.5	\\
 	&	4		&	305.21  &	0.21   	&	-36.8	&	-44.5, -15.9	&	248.0\\
\hline
NW	& 	$2^*$	&	305.41	&	0.25   	&	-32.1	&	-76.3, -32.2	&	2.4 \\
 	&	3		&	305.35  &	0.20   	&	-30.9	&	-87.6, -18.1	&	64.1\\
 	&	5		&	305.35  &	0.15   	&	-32.7	&	-39.9, -20.4	&	294.0\\
\hline
SE	&	8		&	305.33  &	0.07   	&	-42.7	&	-61.3, -26.8	&	217.8\\
 	&	$9^*$	&	305.09  &	0.1		&	-48.6	&	-50.1, -46.6	&	4.8\\
 	&	10		&	305.13  & 	0.08	&	-35.9	&	-34.9, -34.5	&	6.1\\
 	&	11		&	305.20  &	0.00  	&	-36.3	&	-36.8, -24.5	&	12.1\\
 	&	16 		&	305.26	&	-0.1	&	-23.1	&	-22.0, -22.5, 	&	3.6\\
\hline
W	&	6		&	305.72  &	0.09   	&	-39.9	&	-41.8, -38.6	&	8.5\\
 	&	7		&	305.89  &	0.03   	&	-24.5	&	-36.8, -22.2	&	43.6\\	
 	&	$12^*$	&	305.83  &	-0.08  	&	-97.6	&	-120.8, -85.4	&	13.3\\
 	&	$13^*$	&	305.75  &	-0.08  	&	-45.8	&	-45.4, -43.6	&	7.2\\
 	&	$14^*$	&	305.81  &	-0.11  	&	-37.7	&	-38.6, -35.8	&	12.1\\
 	&	15	&		305.80  &	-0.24  	&	-25.4	&	-44.5, -10.4	&	1087.8\\
 \hline
\end{tabular}
\end{table*}

\subsection{Analysis}
In the following section, we present a standard analysis of the ammonia clumps using the spectra shown in Fig. 3.

\subsubsection{Optical depth}

Using the hyperfine structure, we extract the optical depth from the ratio of satellite to main line intensity, using Eq. (1) from \cite{Ho1983}:

\begin{equation}
\frac{\triangle{T^*_\alpha}(\rm J,K,m)}{{\triangle{T^*_\alpha}(\rm J,K,s)}}  = \frac{1-e^{-\tau(\rm J,K,m)}}{1 - e^{-\alpha\tau(\rm J,K,m)}}	
\end{equation}

\noindent where $T^*_\alpha(\rm J,K,m)$ is the observed antenna brightness temperature, (m,s) refer to the main and satellite hyperfine components, $\tau(\rm J,K,m)$ is the optical depth of the main component and $\alpha$ is the ratio of intensity for the satellite compared to the main component [a = 0.28 and 0.22 for the (1,1) satellites]. An implicit assumption is that the beam filling factors are equal as are the excitation temperatures for the different hyperfine components. This is a reasonable assumption because of the very close energy separations and the small probability of special excitation mechanisms that differentiate between the hyperfine components \citep{Ho1983}. 

\subsubsection{Excitation Temperature}

Once the optical depth of the \NH\ line is known the excitation temperature T$_{\rm ex}$ of the \NH\ (1,1) inversion can be derived using Eq. 2 in \cite{Ho1983}:

\begin{equation}
\rm T_{B}(\nu) = T_{0}\left(\frac{1}{e^{T_0/T_{\rm ex}}-1}-\frac{1}{e^{T_{0}/2.7}-1}\right)(1-e^{(-\tau \nu)})
\end{equation}

\noindent where $T_{0}=h \nu/k$, $\rm T_{MB}$ is the main beam telescope brightness temperature and T$_{\rm ex}$ is the source excitation temperature. We assume a background temperature of 2.7 K. This approach yields very low excitation temperatures T$_{\rm ex}$ $\sim3$ K for a beam filling factor of 1, which is a result of beam dilution and smoothing due to the large Mopra beam. This low value of T$_{\rm ex}$ introduces significant error due to being so close to the background temperature and leads to lower limits for the mass of the detected clumps. We therefore assume a more representative excitation temperature of T$_{\rm ex}=10$ K for the rest of our analysis \citep{Harju1993}. We can make an estimate of the excitation by assuming a beam filling factor, it is likely that the beam filling factor is $<10\%$, for a beam filling factor of 0.1 we find the average excitation temperature to be $\sim 7.4$ K, close to our assumed value of T$_{\rm ex} = 10$ K. 

\subsubsection{Column Density}

The column density, $\rm N_u$, of the molecules in the upper state of the transition $u \rightarrow l$ can be expressed as a function of the integrated optical depth of the line using Eq. (3) from \cite{Harju1993}: 
\begin{equation}
\rm N_u = \frac{3h\epsilon_0}{2\pi^2} \frac{1}{|\mu_{ul}|^2} F(T_{\rm ex}) \int  \tau \Delta V 
\end{equation}

\noindent where $|\mu_{ul}|$ is the transition dipole moment (in Cm) and
the function F$_{\rm T_{ex}}$ is defined by: 

\begin{equation}
\rm F(T_{\rm ex}) = \frac{1}{e^{T_0/T_{\rm ex}}-1} 
\end{equation}
\noindent where $T_0 = h\nu_0/k$, the integration of $\tau$ is with respect to 
velocity. The application of the formula above depends on whether we have been
able to determine $\tau$, e.g., from a hyperfine fit as in the case of the (1,1) transition.

\begin{equation}
 \int \, \tau \, dv  \; = \; \frac{\sqrt{\pi}}{2\sqrt{ln 2}} \, 
                          \Delta V \, \tau_0 \;  
\end{equation}
\noindent where $\Delta V$ is the FWHM of the line profile. For the inversion transitions of NH$_3(\rm J,K)$ the squares of the transition dipole moments can be written as:

\begin{equation}
|\mu_{\rm ul}|^2 \; = \; \frac{\rm K^2}{\rm J(J+1)} \, \mu^2 \; 
\end{equation}

\noindent The column density $\rm N_u$ refers to the upper transition level, we make use of the Boltzmann equation with the assumed $\rm T_{ex}$ to estimate the total column density $\rm N(1,1)$ assuming that both levels are evenly populated:

\begin{equation}
\rm N(1,1)=N_{\rm u}+N_{\rm l}=N_{\rm u}(1+e^{hv/kT_{\rm ex}})
\end{equation}

\noindent the metastable inversion transitions, \ie J = K gives the most reliable estimate of the rotational temperature T$_{12}$. Using the (1,1) and (2,2) excitation levels, we make use of the rotational temperature equation from \cite{Wilson1993}:

\begin{equation}
\rm T_{12}=\frac{41.2}{\ln \left [ 3.57\left ( T_{11}/T_{22} \right ) \right ]}
\end{equation} 

\noindent where T$_{11}$ and T$_{22}$ is the main beam brightness temperatures of the \NH\ (1,1) and (2,2) transition. We do not detect \NH\ (2,2) emission in clumps 10, 11 and 15, and therefore we are unable to calculate the rotational temperature for these clumps. However, we are able to calculate a range of values for each parameter of interest such as H$_{2}$ column and number densities and clump masses, using a range of rotational temperatures typical of similar clumps (10 - 40 K; e.g., Wu et al. 2006, Tieftrunk et al. 1998). This range of temperatures corresponds to at most a factor of two difference in column density and so will not affect our results considerably. In order to obtain H$_{2}$ column densities we have assumed an ammonia fractional abundance of 3 \x\ 10$^{-8}$ \citep{Wu2006, Harju1993}.  The H$_{2}$ densities and clump masses have been calculated using the clump area defined by the {\sc Fellwalker} algorithm.  We have made use of the metastable transitions and this may not resemble a true Boltzmann distribution, nevertheless, T$_{12}$ is a good indicator of the gas temperature T$_{\rm kin}$ for temperatures less than 20 K, although T$_{12}$ underestimates T$_{\rm kin}$ for temperature over 20 K (see e.g. \citealt{Walmsley1983, Danby1988}). We are able to calculate the kinetic temperature using: 

\begin{equation}
\rm
T_{\rm kin}=T_{12}\times \left(1+\frac{T_{kin}}{41.5}\right)\times \ln \left ( 1+0.82\times e^{\left ( -21.45/T_{kin} \right )} \right )
\end{equation}

\noindent We estimate the total column density of \NH\ within the clumps using the following equation assuming that only metastable levels are populated:

\begin{equation}
\rm N(NH_{3})=N(1,1)\left(\frac{1}{3}e^{23.4/T_{12}}+1+\frac{5}{3}e^{-41.5/T_{12}}+\frac{14}{3}e^{-101.5/T_{12}}\right)
\end{equation}

\noindent We can test the stability of the cores against collapse by calculating the virial mass, which we derive using the standard equation (e.g., Evans 1999):
\begin{equation}
\rm M_{vir}\simeq210R_{\rm core}\langle\ \Delta V^{2} \rangle
\end{equation}

\noindent where $\rm R_{core}$ is the ammonia clump radius (geometric average dimensions given in Table 2) in parsecs (assuming distance of 4 kpc) and $\Delta V$ is the FWHM line width. The resulting parameters for each clump are presented in Table 6.

\begin{table*}
\caption{\NH\ Clump parameters including; main line and integrated optical depth, rotational temperature, kinetic temperature H$_{2}$ column density, derived mass and virial mass.}
  \begin{tabular}[h]{ccccccccc}
   \hline Region & Clump No & $\btau_{(1,1,m)}$ &$\int \btau_{(1,1)}$& T$_{12}$ & T$_{\rm kin}$ & N(H$_{2}$) & Mass & Virial mass \\
    & & &  & (K) & (K) & 10$^{22}$cm$^{-2}$ & 10$^{3}$(\msun) & 10$^{3}$(\msun) \\
 	\hline
 
   	NE	&	   1    &     0.79    &	6.5	&      21.3  $\pm$   0.6    &
      25.2   $\pm$  0.9    &     6.4   $\pm$   1.8    &     88	$\pm$	
      25    &      31	\\
   		&	3    &      1.13    &	11.3	&      22.3    $\pm$  0.9    &
      26.8   $\pm$   1.5    &      11.2    $\pm$  4.7    &      53	$\pm$
      22    &      24	\\
       \hline
	NW	&     4    &      1.32    &	15.8	&     23.5 $\pm$   1.2    &
      28.8   $\pm$   2.1    &      15.3  $\pm$   7.4    &      122	$\pm$
      59    &      51	\\
   		&    7    &     0.86    & 	5.7	&     20.1  $\pm$   0.8    &
      23.2   $\pm$   1.2    &      5.7    $\pm$  2.7    &      43	$\pm$
      20    &      14 \\
   		&    9    &     0.29    &	1.9	&      19.5    $\pm$  1.1    &
      22.4   $\pm$   1.6    &      1.9   $\pm$   1.2    &      5	$\pm$
      3    &      16	\\
       \hline
	SE	&    2    &     0.79    &	8.3	&      21.1  $\pm$   0.5    &
      24.8   $\pm$   1.2    &      8.2  $\pm$    3.2    &      70	$\pm$
      27    &      37	\\
   		&    8    &      1.03    &	15.0	&      24.4   $\pm$   1.6    &
      30.6   $\pm$   2.9    &      14.6   $\pm$   9.7    &      69	$\pm$
      46    &      54	\\
   		&     6    &     0.47    &	3.5	&      20.6   $\pm$  0.8   &
      24.0   $\pm$   1.3    &      3.5   $\pm$   2.3    &      25	$\pm$
      16    &      17	\\
       \hline
  	W		&   5    &      1.06    &	6.8	&      18.7  $\pm$   0.9    &
      21.2   $\pm$   1.3    &      6.9   $\pm$   3.8   &      46	$\pm$
      25    &      11 \\
  		&   10    &      1.10    & 8.5		&    10 - 40	&
      10 - 74	&      9 - 16        &      21 - 37	&      8	\\
  		&   11    &      1.46    &	10.2	&     10 - 40	&
      10 - 74	&      11 - 20    &      27 - 48    &      9	\\
   		&   12    &     0.36    &	1.8	&      18.3   $\pm$   1.2   &
      20.7  $\pm$    1.8    &      2.0   $\pm$   1.5    &      3	$\pm$
      2    &      2	\\
     	& 	13    &      1.34    & 	11.7	&     20.9  $\pm$    1.2   &
      24.6  $\pm$    1.9   &      11.6   $\pm$   7.3    &      43	$\pm$
      27    &      14	\\
     	& 	 14    &      1.23	&	10.7	&      21.2  $\pm$    1.3    &
      25.1  $\pm$    2.1    &      10.5   $\pm$   7.2    &      22	$\pm$
      15    &      12	\\
      	&	15    &      1.51    &	8.5	&      10 - 40	&
      10 - 74	&    9 - 17    &     11 - 20    &      4	\\
       
\hline

\end{tabular}
\end{table*}

\section{Discussion}

\subsection{General physical properties of the molecular material}
The spatial distribution and morphology of the detected ammonia clumps are given in section 3.1, here we describe the basic parameters. We find \NH\ column densities on the order of $10^{14}$ \cmtwo\ the highest values are found closest to the PDR boundary. Unfortunately we are unable to derive the gas density directly from our measurements, as the derivation of the line excitation temperature which is sensitive to density, requires an assumption of the unknown beam filling factor. We find kinetic temperatures between 21 - 31 K with an average of 25 K, this value is above the ÒHighÓ (22.4 K) and ÒLowÓ (21.7 K) groups of \cite{Molinari1996} and similar to the values found in a survey of ammonia clumps in the Orion and Cepheus clouds by \cite{Harju1993}. An important caveat is that these regions are at different distances to G305 and observed with higher resolution and depth. We assume that our derived column densities are a lower limit due to beam dilution, and the unknown beam filling factor. The \NH\ (1,1) and (2,2) emission appears to be biased to cool gas which is extended over the size of the Mopra beam and does not reflect the physical conditions of the unresolved dense cores and filaments we expect to find at sub arc minute scales. The presence of hotter gas is indicated by (3,3) emission and so clumps 1 - 5 and 8 are likely to be more evolved than other, cooler clumps. The clumps that exhibit (3,3) emission are found to correlate well with PAH emission, with the strongest and therefore hottest clumps found close to the central cavity boundary.

Using an assumed \NH\ to H$_{2}$ abundance ratio of $3\times10^{-8}$ we calculate H$_{2}$ column densities of $>10^{22}$ \cmtwo. We assume that the ammonia abundance fraction is constant across the entire region, which is not likely to be the case, however with our current low resolution observations our chosen value gives an appropriate estimate. The abundance ratio of \NH\ to H$_{2}$ has been found to vary from $1\times10^{-7}$ to $1\times10^{-8}$ according to the study of Orion clumps \citep{Harju1993}, thus our measurements of the H$_{2}$ column density could be an order of magnitude lower than shown in Table 6 and so our results should be taken as an upper limit. 

The sizes of the \NH\ clumps have been derived using the geometric mean of the linear x and y distances, they range from 2.6 to 10.1 pc. The sizes of the (2,2) emission are smaller than those of the (1,1) emission in all cases and in the same way (3,3) emission is smaller than (2,2) emission. The size of these clumps is significantly greater than those detected in a survey carried out by \cite{Wu2006} (0.4-3.1 pc) and similar observation in the Orion and Cepheus cloud carried out by \cite{Harju1993} (mean 0.3 pc). However the same differing resolution and sensitivity caveat applies, and due to our large beam we are only able to resolve \NH\ structure of $>2.6$ pc.

We calculate masses for the clumps in excess of a thousand solar masses, with a combined mass of $\sim6\times10^{5}$ \msun. It is clear that there is sufficient dense molecular gas to form multiple massive stellar clusters within the detected clumps, and this is likely to occur around the central region. The virial mass is found to be significantly less than the calculated mass for all clumps, we would therefore assume these clumps are gravitationally bound and given the large difference between their actual masses and their virial masses it is likely they are in a state of gravitational collapse, however this may not be an appropriate comparison. The mass of the detected clumps in G305 is higher than expected for isolated clumps that tend to have mass on the order of a few tens to hundreds of solar masses. The simplest explanation for this is that our detected clumps are in fact comprised of a number of unresolved clumps. In previous higher resolution studies of \NH\ clumps \cite{Harju1993} the average size of the clumps is significantly lower than our resolving limit of 2.6 pc, therefore it is no surprise that the average mass of these smaller clumps in the Cepheus, Orion and Taurus clouds are found to be 15 or 140, 15 and 4 \msun\ significantly less than our estimates. Given these average clump masses and sizes and comparing to our derived values it is apparent that the  clumps that we have detected in G305 most likely consist of a significant number of unresolved clumps and therefore may be better described as`` clouds'' rather than ``clumps''.

As mentioned above we are unable to directly estimate the gas density however we can make estimate of the density of the ammonia clumps by assuming the clumps are spherically symmetric and our estimation of the excitation temperature is correct. This results in a  H$_2$ number density of $\sim1\times10^3$\cmthree for the largest clumps in the NE, NE and SE region, significantly lower than the critical density of ammonia of  $10^4 - 10^5$ \cmthree, we find that the smallest clumps detected in the western region have the highest H$_2$ number densities of $\sim10^4$. This is further evidence that the large clumps we find in the NE, NW and SE region are in fact comprised of unresolved sub structure which are smoothed by the beam into the single clumps we detect. It appears that the mass of several unresolved clumps is spread out over the detected area by the $\sim 2 \arcmin$ beam, this drives down the estimated density for the largest clumps but has less of an effect on the smaller clumps which are likely to consist of at most a few cores. This means that the mass we estimate for the largest clumps is most likely an overestimate as the area of emission appears greater than it physically is.

There is compelling evidence of substructure within the detected clumps, this coupled with the unknown abundance ratio, contribute the main sources of error in our calculations. Thus our results should be taken as an upper limit due to the unknown beam filling factor and may be up to an order of magnitude too high due to the unknown abundance ratio.

\subsection{Star formation and ammonia clumps towards G305}

A key goal of our observations was to study the relationship between the dense molecular gas in G305, as traced by \NH\ and the distribution of known sites of recent or ongoing star formation. To locate areas of ongoing star formation we make use of the Red Source MSX Survey (RMS) \citep{Urquhart2008,Hoare2005} and 6.7 GHz methanol masers drawn from \cite{Caswell2009} and the  Methanol Multi Beam (MMB) survey \citep{Green2009}, as well as our \water\ maser results. In Figures 5 - 8, we present a panoramic view of the ongoing star formation towards G305.

The RMS survey aims to locate massive young stellar objects (MYSOs), and so highlight regions of recent star formation. Colour selection in the MSX point source catalogue is used to identify objects with colours the same as well known MYSOs. Many objects have very similar colours to the very red MYSOs, this includes UC \HII\ regions, compact planetary nebula, low mass YSOs and evolved stars. We have selected objects that are either MYSOs, YSOs, \HII\ or UC \HII\ detections. We do not distinguish between the different types of selected RMS sources as we will be using them only as a general indicator of recent star formation and have insufficient numbers to carry out any statistical analysis. We find 26 RMS sources in the direction of G305 and overplot these with red boxes in Figures 5 - 8. 

The Methanol Multi Beam survey \cite{Green2009} is the widest area survey yet carried out for 6.7 GHz methanol masers, comprising an initial maser search with Parkes and sub-arcsecond positional followup by the Australia Telescope Compact Array and MERLIN. Methanol masers have been recognised as one of the brightest signposts to the formation of massive young stars, \citep{Menten1991} and unlike other strong masers (OH, \water\, and SiO) have thus far only found close to high mass star forming regions \citep{Minier2003}. The MMB survey has not yet published their catalogue for the $l$=305 region and so we use preliminary maser positions kindly provided by the MMB team (Fuller \& Caswell, priv.~comm.) and masers previously documented by \cite{Caswell2009} we  identify 13 6.7 GHz methanol masers in the direction of G305 and overplot their locations with cyan circles in Figures 5 -- 8.  
Below we briefly describe the locations of star formation tracers with respect to the ammonia emission.\\

\begin{figure}
\caption{\NH\ (1,1) and (2,2) emission contoured over greyscale GLIMPSE 5.4\,$\umu$m image in black and red respectively clump numbers are shown within circles. Detected \water\ masers are shown by blue crosses, MMB sources by yellow circles and RMS sources by red boxes. Contours begin at 0.15 K and increment by 0.1 K.}
\includegraphics[trim= 30 0 0 0, width=0.5\textwidth]{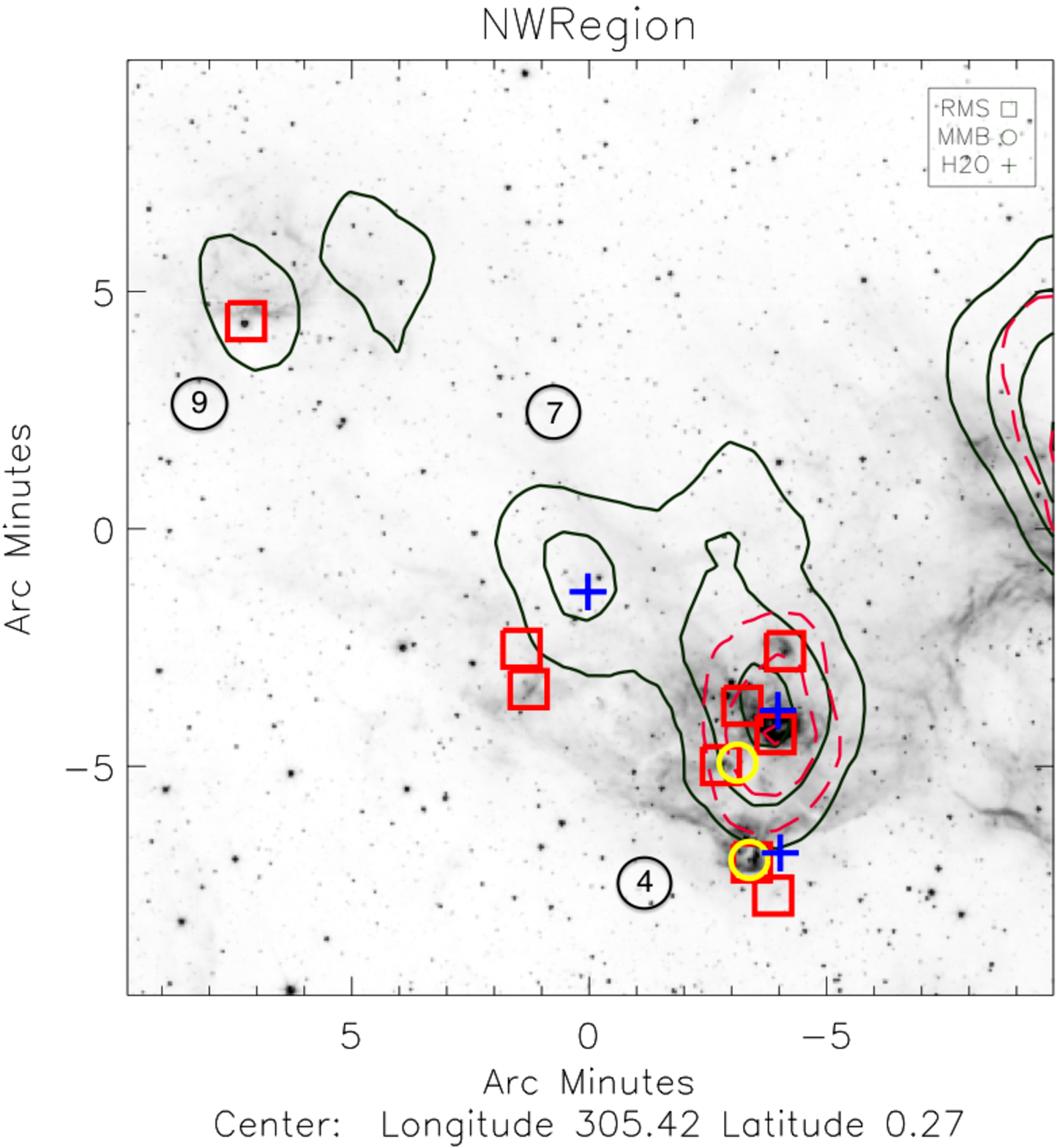}
\label{NW Region}
\end{figure}

\noindent \textbf{NW Region:} In this region (Fig. 5) we note the highest number of RMS sources (nine) of which six are found within and around the periphery of clump 4. Two RMS sources are seen to the west of clump 7 and one RMS source is seen within clump 9. We find two methanol masers associated with clump 4, one is seen just outside the leading edge of the ammonia emission facing WR48 a, the other methanol maser is seen towards the centre of the clump. Three \water\ masers are detected, two of which reside within clump 4 and one within clump 7 again close the clump centres.
It is interesting that we see RMS sources, \water\ masers and methanol masers all within the bounds of clump 4. as mentioned in section 3.1 unlike other regions we see that the clump is coincident with strong PAH emission towards the centre of clump. High resolution observations are required to resolve the internal structure and kinematics but it is likely that there are a number of individual dense cores at different velocities contributing to the large FWHM line width of clump 4. The RMS and MMB sources appear to lie around the core of the detected clumps and are predominantly found on the western side of the clump. The \water\ masers again are located close to the clump cores. Nine 1.2 mm cores are detected by \cite{Hill2005} six are found within clump 4, three are found in clump 9.
Clump 1 in the NE region is the largest clump in geometric size but clump 4 is the most massive\\

\begin{figure}
\caption{\NH\ (1,1) and (2,2) emission contoured over greyscale GLIMPSE 5.4\,$\umu$m image in black and red respectively, clump numbers are shown within circles. Detected \water\ masers are shown by blue crosses, MMB sources by yellow circles and RMS sources by red boxes. Contours begin at 0.15 K and increment by 0.1 K.}
\includegraphics[trim= 30 0 0 0 ,width=0.5\textwidth]{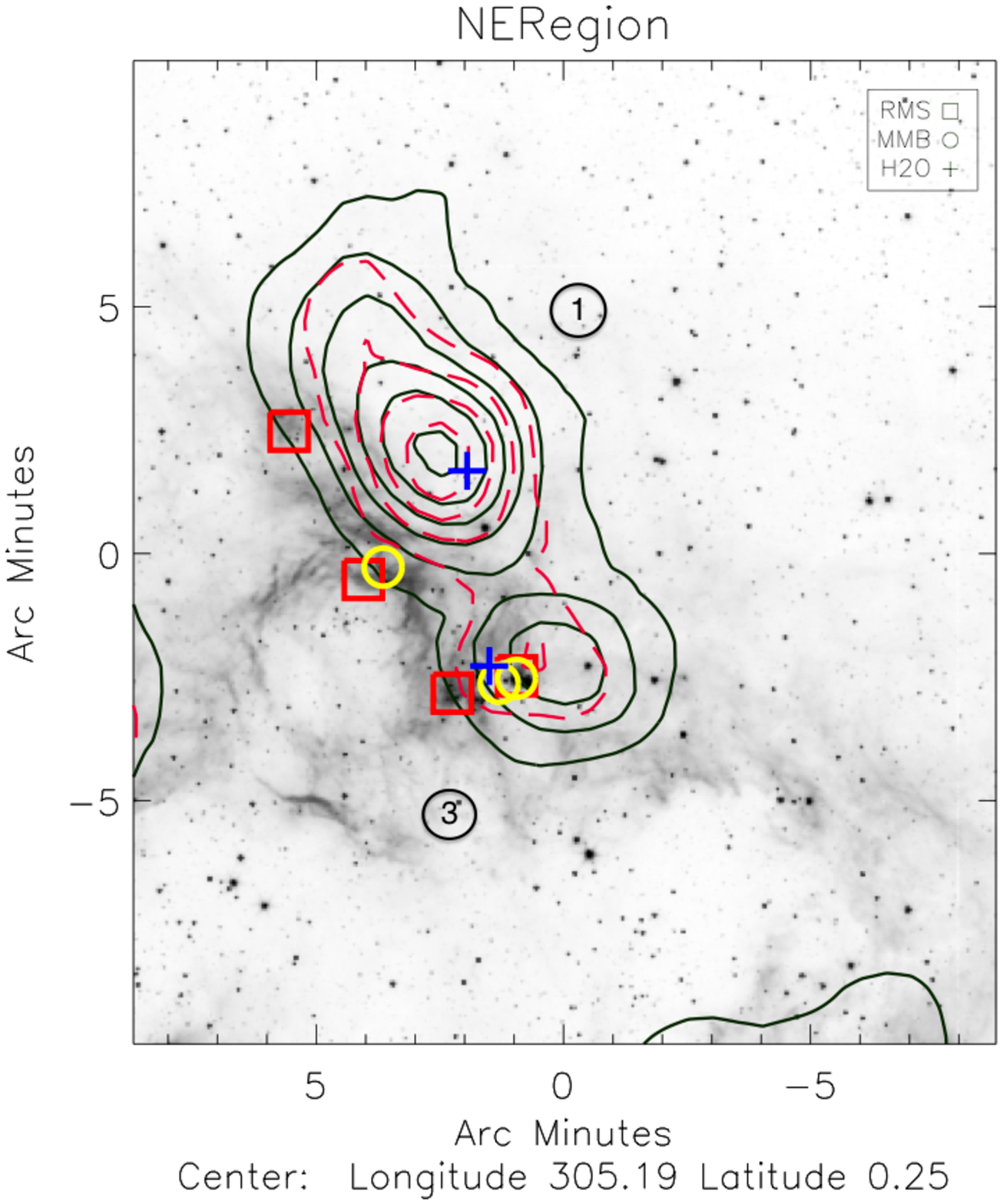}
\label{NE Region}
\end{figure} 

\noindent \textbf{NE Region:} In the bounds of this region (Fig. 6) we find four RMS sources, three of which are found along the western edge of the \NH\ emission, one is located close the centre of clump 3. There are three methanol masers, two of which (G305A and B) are found close the centre of clump 3 and one on the western edge of clump 1 coincident with an RMS source and strong PAH emission. The two methanol masers G305A and B are found to be coincident with CH$_{3}$OH emission, which surrounds the two masers and is elongated to the north. We note the position of a \HII\ region (see \cite{Clark2004}) to the west of the detected \NH\ clumps, seen as a ring in the 5.4\,$\umu$m PAH emission, this coincides with the low density region separating clump 1 and 3. Star formation tracers are clustered on the western side of the region facing this \HII\ bubble, there are no signs of star formation to the east of the clumps. We detect two \water\ masers towards the centre of two clumps. It appears that RMS and MMB star formation tracers are confined within the lower density \NH\ facing the ionising \HII\ region to the west, the \water\ masers however are found embedded towards the centre of the two clumps.

This region has been studied in some detail \citep{Walsh2006,Hill2005,Longmore2007}. 
\cite{Hill2005} detect nine 1.2 mm cores within this region concentrated within clump 1 but offset from the peak to the west. Observations in the mid infrared \cite{Walsh2001}, near infrared \cite{Walsh1999} and  mm line observations \cite{Walsh2006} ($^{13}$CO, HCO$^{+}$, N$_{2}$H$^{+}$, CH$_{3}$CN and CH$_{3}$OH) were carried out towards the two methanol maser sites designated G305A and G305B (G305.21+0.21 and G305.20+0.21). These methanol masers are shown in Fig. 6 as the most southern two masers, the centre of the detected molecular emission was found to be G305A which has no detected infrared or radio continuum (however we do find it is coincident with a 1.2 mm core). Walsh concludes that G305A is a site of massive star formation similar to the Class 0 stage of low-mass star formation. G305B is coincident with a bright, red infrared source whose bolometric luminosity suggests it should produce a detectable \HII\ region, yet none has been detected. This suggests that G305B may also be at a very early stage of evolution, but probably older than G305A. Longmore detect twelve embedded sources with significant IR excess in this region located around the southern edge of clump 3 within the region of low emission separating clump 1 and 3. An embedded source is found in the centre of the \HII\ region highlighted by 5.4\,$\umu$m PAH emission to the west of clump 1 and 3. \cite{Longmore2007} conclude that the expanding bubble may be responsible for triggering a third generation of star formation in clump 1 and 3. \\

\begin{figure}
\caption{\NH\ (1,1) and (2,2) emission contoured over greyscale GLIMPSE 5.4\,$\umu$m image in black and red respectively, clump numbers are shown within circles. Detected \water\ masers are shown by blue crosses, MMB sources by yellow circles and RMS sources by red boxes. Contours begin at 0.15 K and increment by 0.1 K.}
\includegraphics[width=0.5\textwidth]{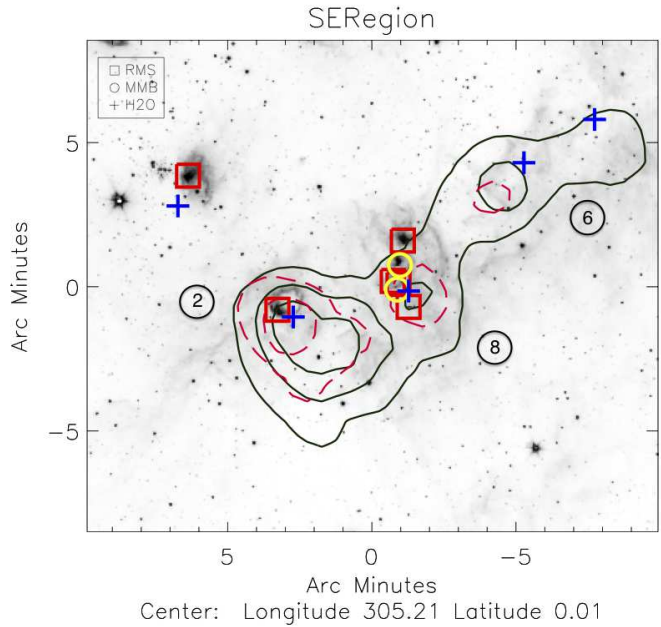}
\label{SE Region}
\end{figure}

\noindent \textbf{SE Region:} This region is shown in Fig. 7, coincident with this strip of \NH\ emission we find five RMS sources, four \water\ masers and two methanol masers. We also note that in the central cavity of G305 there is an RMS source and \water\ maser coincident with strong PAH emission but devoid of any \NH\ emission. The Four \water\ masers within the \NH\ emission are all in close proximity to the cores of the \NH\ emission. A striking feature of the region is that four of the five RMS sources and both methanol masers are clustered within the low density region of clump 8 and extend towards the core of the clump, this area is also the location of a \HII\ region documented in \cite{Clark2004}, this is clearly a very active area of ongoing star formation. All of these star formation tracers are found on the inward side of the ammonia emission close to the boundary between the central cavity, again facing the ionising central sources. \cite{Hill2005} detect seven 1.2 mm cores, four of these are found in the intense star forming region between clumps 8 and 6.\\
 
\noindent \textbf{W Region:} This region, shown in Fig. 8 is distinctly different from the other regions we have discussed. We find a total of eight RMS sources, six MMB sources and six \water\ masers. Whereas the previous regions all show strong \NH\ emission around the periphery of the central \HII\ region, this region shows no significant \NH\ at this boundary. However, in the absence of any significant ammonia emission at the boundary we do still see significant signs of star formation coincident with clump 14 close to the cavity boundary, with four RMS sources and a single MMB source. Further to the west of the cavity we see dispersed signs of ongoing star formation the majority of which are associated with isolated \NH\ emission. Clump 5 is host to three \water\ masers, the most detections coincident with a clump in the complex. This region is also host to the most methanol maser detections, 5 of which are found on the edge of \NH\ clumps. \cite{Hill2005} detect eleven 1.2 mm cores, which are confined to clump 14 and clump 13. Unlike the other regions where 1.2 mm emission is correlated with strong \NH\ (1,1) and (2,2) emission, there is no 1.2 mm emission associated with the strongest ammonia emission in this region, in clump 5. It is clear that this region is host to star formation, however there is much less molecular material in this region and the star formation appears much more dispersed. \\

\begin{figure}
\caption{\NH\ (1,1) and (2,2) emission contoured over greyscale GLIMPSE 5.4\,$\umu$m image in black and red respectively, clump numbers are shown within circles. Detected \water\ masers are shown by blue crosses, MMB sources by purple circles and RMS sources by red boxes. Contours begin at 0.15 K and increment by 0.1 K.}
\includegraphics[ width=0.5\textwidth]{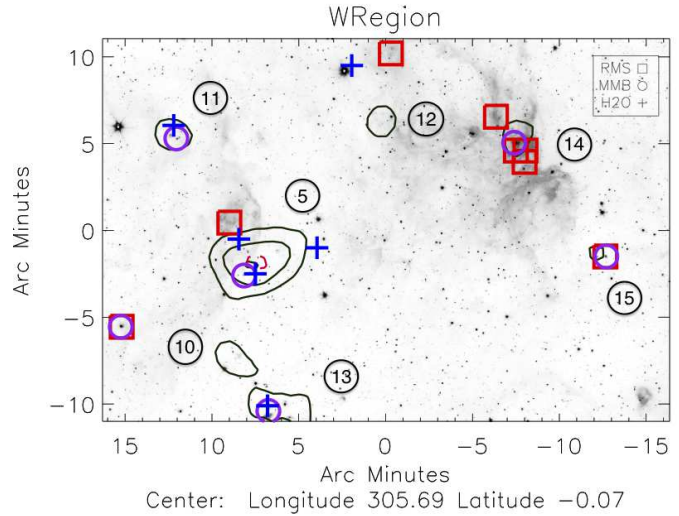}
\label{W Region}
\end{figure} 

It is clear that all regions are host to significant signs of ongoing star formation. Of all the detected clumps only clump 1 and 3 have been studied in any detail by the authors (mentioned in the NE clump section). A striking feature of all these signs of star formation is that they are for the most part all coincident with dense molecular material and  are located only on the side of the molecular emission facing the PDR. In cases where we appear have a side on line of site we see star formation only occurring in the low density regions of the ammonia clumps with exception to \water\ masers, which appear closer to the dense clump cores. This is most apparent in the NE and S regions, where RMS and MMB detections are seen only in the less dense gas towards the PDR, whereas \water\ masers are found deeply embedded in the \NH\ emission. This may indicate an age spread in the star formation with \water\ masers highlighting a younger age of star formation embedded towards the dense clump centres and older YSOs and visible \HII\ regions found on the clump periphery in the lower density regions. Warmer clumps, highlighted by \NH\ (3,3) emission, correlate well with regions of intense star formation, indeed all clumps with (3,3) detections are either host to a large number of star formation tracers or strong PAH emission.

In terms of the star formation distributed within G305 and its relation to the dense molecular material surrounding the central cavity, we have expanded on the work of \cite{Clark2004}, who suggested that the morphology of the nebula is strongly suggestive of star formation on the periphery of the cavities   (see \citealt{Clark2004} Section 4.3). We confirm that star formation is taking place on the periphery of the central cavity created by Danks 1 \& 2, WR48 and also that a potentially younger generation of star formation is taking place around the younger \HII\ regions located around the central cavity, indicating multiple epochs of massive star formation in G305. The dense molecular material surrounding Danks 1 \& 2 takes the form of several discrete and massive clumps, with significant ``gaps''  (particularly towards the western region) highlighting areas where the ionising photons from these massive stellar clusters  can escape. Thus the integrated radio flux very likely underestimates the true ionising photon flux from G305.
We are still unable to ascertain if the photoionisation driven shocks are responsible for triggering the observed star formation, however we do note a possible age spread in star formation tracers in the regions closest to the cavity. The western region may be more likely to be spontaneous in nature star formation due to its dispersed morphology.

\subsection{Comparison}
\begin{table}
\caption{G305 Comparison, column 3 is molecular mass and column 5 shows the number of massive OB stars in the complex. Parameters have been taken from a literature search, numbers in brackets correspond to the following papers.}
  \begin{tabular}[h]{ccccc}
   \hline Region & Distance & Mass & NLyc & Stars\\
   			& kpc	& \msun	& $10^{49}$ s$^{-1}$	& OB \\
   	\hline
   	
   	G305			&	3.5 - 4	(1)		&	$10^{5}$ (*)	&	145.5 (2)	&	$>31$ (1)	\\
   	W49A			&	10.2 - 12.6 (4)	&	$10^{6}$ (3)	&	171.7 (2)	& $>40$ (4)	\\
   	NGC3603			&	5 - 8 (5)		&	$10^{5}$ (5)	&	187.7 (2)	& 	20 (6) 		\\
   	M17				&	1.3 - 1.9 (9)	&	$10^{4}$ (9)	&	53.7 (2)	&	16 (9)		\\
   	W3 main			&	1.9 - 2.1 (10)	&	$10^{4}$ (10)	&	8.8	(2)		&	12 (10)		\\
   	Rosette Nebula	&	1.4 - 1.6 (11)	&	$10^{5}$ (11)	&	9.8	(14)	&	30 (11)	\\
   	Westerlund 2	&	5 - 7 (12)		&	$10^{5}$ (12)	&	95.9 (2)	&	$>12$ (12)	\\
   	Carina			&	2.3	(13)		&	$10^{5}$ (13)	&	21.6 (2)	&	65 (13)		\\

 \hline
 \newline
\end{tabular}
{\footnotesize (*) This work (1) \cite{Clark2004} (2) \cite{Smith1978} (3) \cite{Smith2009} (4) \cite{Depree1997} (5) \cite{Nurn2002} (6) \cite{Moffat1983}  (7) \cite{Dough2009} (8) \cite{Luna2009} (9) \cite{Povich2009} (10) \cite{Tief1998} (11) \cite{Wang2008}  (12) \cite{Dame2007} (13) \cite{Smith2008} (14) \cite{Churchwell1975}}
\end{table}

To put G305 into context it is useful to compare the region to other well known GMCs and \HII\ regions in the galaxy. We have performed a literature search and present parameters for comparison in Table 7. It is apparent that G305 is one of the closest and most massive GMCs in the galaxy, with a molecular mass approximately equal to or exceeding that of other well studied regions. G305 is significantly closer than the regions with similar masses detailed in Table 7, and regions that are found closer to the sun tend to have a lower molecular mass. With its numerous \HII\ regions, the integrated Lyc photon count is on par with the most luminous \HII\ regions in the galaxy. However, whereas many of these regions comprise single \HII\ regions, G305 is host to a number of individual \HII\ regions surrounding a central cavity. As proposed in \cite{Clark2004} given the morphology of the G305 complex and in particular the fact that sufficient natal material has been dispersed to allow Danks 1 and 2 to become optically visible, there is a likelihood of substantial photon leakage and thus the massive stellar population of G305 is likely to be an underestimate. This would make G305 one of the closest examples of intense massive star formation in the galaxy, comparable to W49A which lies at $\sim 3-4$ times the distance.

\subsection{Summary}

We present radio observations of the $\sim1.5\degr \times 1\degr$ region G305 in order to uncover the dense molecular component of the region. This study includes observations of the \NH\ (1,1), (2,2) and (3,3) spectral lines, as well as the 22 GHz \water\ maser line. To complement our observations we have made use of mid-IR data from the GLIMPSE survey and traced signs of star formation using the RMS and MMB surveys. Combining these data sets has allowed us to obtain a panoramic view of ongoing star formation within G305 and its relation to the dense molecular material.

We detect 15 \NH\ (1,1) clumps, 12 \NH\ (2,2), 7 \NH\ (3,3) clumps and 16 \water\ masers. Our observations reveal that G305 is comprised of several massive molecular clumps mostly found towards the northern and southern edges of the main cavity. The molecular material towards the western region of G305 is more dispersed in nature. Kinetic temperatures of the clumps ranges from $\sim21 - 31$ K, the total mass of dense gas traced by \NH\ is estimated to be $\sim6\times10^{5}$ \msun\ with an average clump mass of $\sim4\times10^{4}$ \msun. We detect the majority ($\sim$80\%) of this material in close proximity to the PDR surrounding the central cavity. It is clear that there is sufficient material to form a number of massive clusters and that there is a high degree of interaction between the massive stars and the molecular material. It is possible that such interactions could be responsible for triggering the star formation seen around the periphery of the \NH\ clumps. 

We find 27 RMS sources and 13 MMB masers, as well as 16 \water\ masers. With 56 star formation tracers G305 is obviously a region of intense star formation. We are able to clearly identify the PDR and find it correlates with the molecular gas traced by \NH\. Star formation appears to be taking place within and on the periphery of the detected \NH\ clumps, with $>80\%$ of star formation tracers found to be associated with \NH\ emission. In almost all cases star formation is located on the side of the \NH\ clump that faces the ionising radiation highlighted by the PDR, evidence perhaps of triggered star formation. 

These low resolution observations are intended as a pathfinder for future high resolution follow up, which will allow us to uncover the sub parsec structure of the ammonia emission we predict is present, and thus allow us to comment on the  effects massive star formation is having upon the dense molecular material.

\section*{Acknowledgments}
We would like the thank the Director and staff of the Paul Wild Observatory for their assistance with a pleasant and productive observing run. We would like to thank the referee, Michael Burton, for a thoroughly constructive and useful report. The Mopra telescope  is part of the Australia Telescope which is funded by the Commonwealth of Australia for operation as a National Facility managed by CSIRO. The University of New South Wales Digital Filter Bank used for the observations was provided with support from the Australian Research Council. This research has made use of the NASA/ IPAC Infrared Science Archive, which is operated by the Jet Propulsion Laboratory, California Institute of Technology, under contract with the National Aeronautics and Space Administration.

\label{lastpage}

\end{document}